\documentclass[useAMS,usenatbib]{mn2e}
\usepackage{graphicx}
\usepackage{afterpage}

\newcommand{\aap}{A\&A}

\newcommand{\aj}{AJ}
\newcommand{\mnras}{MNRAS}

\newcommand{\apjs}{ApJS}
\newcommand{\MC}{\multicolumn}

\newcommand{\AM}{\mbox{AM1934-563}}
\newcommand{\Na}{Na\,I}
\newcommand{\Mg}{Mg\,I}
\DeclareRobustCommand{\ionn}[2]{%
\relax\ifmmode
\ifx\testbx\f
{\mathbf{#1\,\textsc{#2}}}\else
{\mathrm{#1\,\mathsc{#2}}}\fi
\else\textup{#1\,{\mdseries\textsc{#2}}}%
\fi}

\def\etal{{\it et al.\ }}
\def\eg{{\it e.g.,}\,}

\title[The polar ring galaxy \AM\ revisited]
{The polar ring galaxy \AM\ revisited
\footnotemark[0]\thanks{%
Based on observations obtained with the Southern African Large
Telescope (SALT).} }
\author[Noah Brosch et al.]{%
Noah Brosch,$^{1,2}$\thanks{E-mail: noah@wise.tau.ac.il (NB); akniazev@saao.ac.za (AYK);
dibnob@saao.ac.za (DB); dod@saao.ac.za (DOD); hashimot@saao.ac.za (YH);
nsl@saao.ac.za (NL); erc@saao.ac.za (ER); still@saao.ac.za (MS);
petri@saao.ac.za (PV); ebb@sal.wisc.edu (EBB); khn@sal.wisc.edu (KN)
}
Alexei Y. Kniazev,$^{2,3}$
David Buckley,$^{2}$
Darragh O'Donoghue,$^{2}$
\newauthor
Yas Hashimoto,$^{2}$
Nicola Loaring,$^{2}$
Encarni Romero,$^{2}$
Martin Still,$^{2}$
\newauthor
Petri Vaisanen,$^{2}$
Eric B. Burgh,$^{4}$
Kenneth Nordsieck$^{4}$\\
$^{1}$The Wise Observatory and the School of Physics and
Astronomy, the Raymond and  Beverly Sackler Faculty of Exactâ
Sciences, \\ Tel Aviv University, Tel Aviv 69978, Israel \\
$^{2}$South African Astronomical Observatory, Observatory Road, Cape Town, South
Africa \\
$^{3}$Special Astrophysical Observatory, Nizhnij Arkhyz, Karachai-Circassia,
369167, Russia \\
$^{4}$Space Astronomy Laboratory, University of Wisconsin, Madison, WI 53706, USA
}

\begin{document}

\date{Accepted 2007 April ??. Received 2007 March ??; in original form 2007 March ??}

\pagerange{\pageref{firstpage}--\pageref{lastpage}} \pubyear{2007}

\maketitle

\label{firstpage}


\begin{abstract}
We report long-slit spectroscopic observations of the dust-lane
polar-ring galaxy \AM\ obtained with the Southern African Large
Telescope (SALT) during its performance-verification phase. The
observations target the spectral region of the H$\alpha$, [\ionn{N}{ii}]
and [\ionn{S}{ii}] emission-lines, but show also deep \Na\ stellar
absorption lines that we interpret as produced by stars in
the galaxy. We derive rotation curves along the major axis of the
galaxy that extend out to about 8 kpc from the center for both the
gaseous and the stellar components, using the emission and
absorption lines. We derive similar rotation curves along the
major axis of the polar ring and point out differences between
these and the ones of the main galaxy.

We identify a small diffuse object visible only in H$\alpha$
emission and with a low velocity dispersion as a dwarf \ionn{H}{ii} galaxy
and argue that it is probably  metal-poor. Its velocity indicates
that it is a fourth member of the galaxy group in which \AM\,
belongs.

We discuss the observations in the context of the proposal that
the object is the result of a major merger and point out some
observational discrepancies from this explanation. We argue that
an alternative scenario that could better fit the observations may
be the slow accretion of cold intergalactic gas, focused by a
dense filament of galaxies in which this object is embedded.

Given the pattern of rotation we found, with the asymptotic
rotation of the gas in the ring being slower than that in the disk
while both components have approximately the same extent, we
point out that \AM\ may be a galaxy in which a dark matter halo is
flattened along the galactic disk and the first object in which
this predicted behaviour of polar ring galaxies in dark matter
haloes is fulfilled.

\end{abstract}

\begin{keywords}
galaxies: ring galaxies --- galaxies: evolution
 ---  galaxies: individual: \AM\ --- galaxies: dark matter ---
galaxies: galaxy haloes
\end{keywords}

\section*{Introduction}

Ring galaxies posed significant astronomical interest since
\citet{LT76} modelled the Cartwheel galaxy as the result of a
small galaxy passing through a larger one. While such events
probably happen and produce some of the ring galaxies, in other
instances different mechanisms might be at work. A particularly
interesting kind of ring galaxy is the polar ring galaxy (PRG)
where a flattened disk galaxy exhibits an outer ring of stars and
interstellar matter that rotate in a plane approximately
perpendicular to the central disk. An extensive catalog of PRGs
was produced by \citet{Whietal90}.

The issue of PRGs was reviewed by \citet{Co06}. She reviewed a
number of formation mechanisms for PRGs: minor or major mergers,
tidal accretion events, or direct cold gas accretion from
filaments of the cosmic web. \citet{Co06} proposed that these objects
can be used to probe the three-dimensional shape of dark matter
(DM) haloes, provided the PRG is in equilibrium in the
gravitational potential.

The well-known Spindle Galaxy (NGC 2685), an archetypal PRG,
exhibits two sets of rings: an outer one visible only on HI maps
and which might be in the plane of the galaxy, and an inner one
that is helix-shaped, is perpendicular to the main axis of the
galaxy, is optically bright, shows embedded present-day star
formation, and is associated with prominent dust lanes. Shane (1980)
explained the system as consisting of a lenticular galaxy
that recently accreted an HI gas cloud that formed the inner ring,
while the outer gas ring might be a remnant of the formation of
the galaxy. Hagen-Thorn \etal (2005) found that the stellar
population of the inner system of dust and gas, arranged in a
spiral around the ''spindle'' but really in a disk, is
1.4$\times10^9$ years old.

In a different ring galaxy, NGC 660, Karataeva \etal (2004)
detected red and blue supergiants belonging to the ring system.
They showed that the age of the youngest stars there is only
$\sim$7 Myr; thus star formation is currently taking place. N660
is special in that both the disk and the polar ring contain stars,
gas and dust. \citet{Resh04}, who analyzed three other ring
galaxies, showed that their rings result from ongoing interactions
or mergers where the main galaxy is a spiral and the rings are
currently forming stars.

Other claims of interactions being at the origin of the rings and
of the star formation taking place therein have been put forward
by Mayya \& Korchagin (2001, revised 2006). On the other hand,
others claimed that rings are formed as a dynamical event in a
larger disk galaxy (e.g., Mazzuca \etal 2001). It is clear that
more studies of ring galaxies, and in particular such
investigations that can time the ring and star formation events,
can help understand the particular instances when a galaxy-galaxy
interaction took place, when a ring is formed, and when the event
does trigger the SF process. There is also the possibility that
careful tracing of the polar ring and of the galaxy itself, and
their kinematic properties, might reveal the DM halo shape and
properties, as advocated by \citet{Co06}. This singles out PRGs as
valuable targets for DM studies.

In this paper we analyze new observations of the polar-ring galaxy
\AM, a PRG with an optical redshift of 11649$\pm$10 km sec$^{-1}$
located at l=341.02, b=-28.73, also identified as PRC B-18 in
Whitmore \etal (1990). The object was recently studied by
\citet{Resh06}, who showed that this is a giant galaxy in a
compact triplet, together with PGC~400092 (classified Sd/Irr:) and
PGC~399718 (classified  SBc:) at approximately the same redshift.
The authors used the 1.6-meter telescope of the Pico dos Dias
Observatory in Brazil for imaging in BVRI, the CTIO 1.5-meter
telescope to collect spectral observations, and included data from
IRAS and 21-cm line observations. However, most of their
conclusions about the nature of the object rely on the
morphological appearance of the galaxy.

\citet{Resh06} modelled \AM\ using an N-body code that includes
gas dynamics using sticky particles and star formation. They
concluded that the best-fitting model is of a major merger,
whereby a gas-rich galaxy transferred a sizable amount of matter
to \AM\ during a parabolic encounter. The matter subsequently
relaxed and now forms a complete ring of stars, gas, and dust
around \AM\, whereas the donor galaxy is one of the two other
galaxies in the same group.

The reason to revisit this object was the availability of
high-quality spectra obtained with the effectively 8-meter
diameter Southern African Large Telescope
(SALT) telescope. We derive, for the first time, rotation
curves for the ionized gas and for the stellar components of both
the main galaxy and the polar ring. Since PRGs might make good
test cases for the properties of dark matter haloes in and around
galaxies, as argued by \citet{Co06}, the more observational data
collected on these objects and with higher quality, the better.

Very few PRG observations obtained with large telescopes have been
published. A noticeable one is by Swaters \& Rubin (2003), with
the Baade 6.5-meter telescope on Las Campanas, tracing the
dynamics of the stellar component of the prototype PRG NGC 4650A
where they showed that the polar ring is actually a polar disk, an
extended feature rather than a narrow gas disk. They favour a
scenario by which the ring/disk was formed  by the polar merger to
two similar disks, as previously suggested by Iodice \etal (2002).
Iodice \etal (2006) observed the gaseous component in the ring of
N4650A with ESO's FORS2 on UT4 and concluded that a scenario by
which it could be formed was through slow gas accretion from the
cosmic web filaments. We propose that the same situation could be
taking place for \AM.

This paper is organized as follows: \S~\ref{txt:Obs_and_Red} gives
a description of all the observations and data reduction. In
\S~\ref{txt:results} we present our results, analyze them in
\S~\ref{txt:disc}, and present our interpretation in
\S~\ref{txt:interp}. The conclusions drawn from this study are
summarized in \S~\ref{txt:summ}.


\begin{table}
\caption{Details of the \AM\ RSS observations}
\label{t:Obs}
\begin{tabular}{cccccc} \hline
  Date        & Exp.time     & Spec. Range    & Slit    &  \MC{1}{c}{PA} & Disp.        \\
          &  (sec)       &       (\AA)    &(\arcsec)& ($^\circ$)     & (\AA/pix)    \\ \hline
  16.07.2006  & 2$\times$600 & 3650--6740     &  1.5    &  140           & 0.98          \\
  16.07.2006  & 1$\times$600 & 3650--6740     &  1.5    &  35            & 0.98          \\
  20.09.2006  & 2$\times$900 & 6050--7315     &  1.5    &  140           & 0.40          \\
  20.09.2006  & 1$\times$750 & 6050--7315     &  1.5    &  27            & 0.40          \\
  21.09.2006  & 3$\times$900 & 6050--7315     &  1.5    &  27            & 0.40          \\  \hline
\end{tabular}
\end{table}

\begin{figure*}
{\centering
 \includegraphics[clip=,angle=0,width=15.0cm]{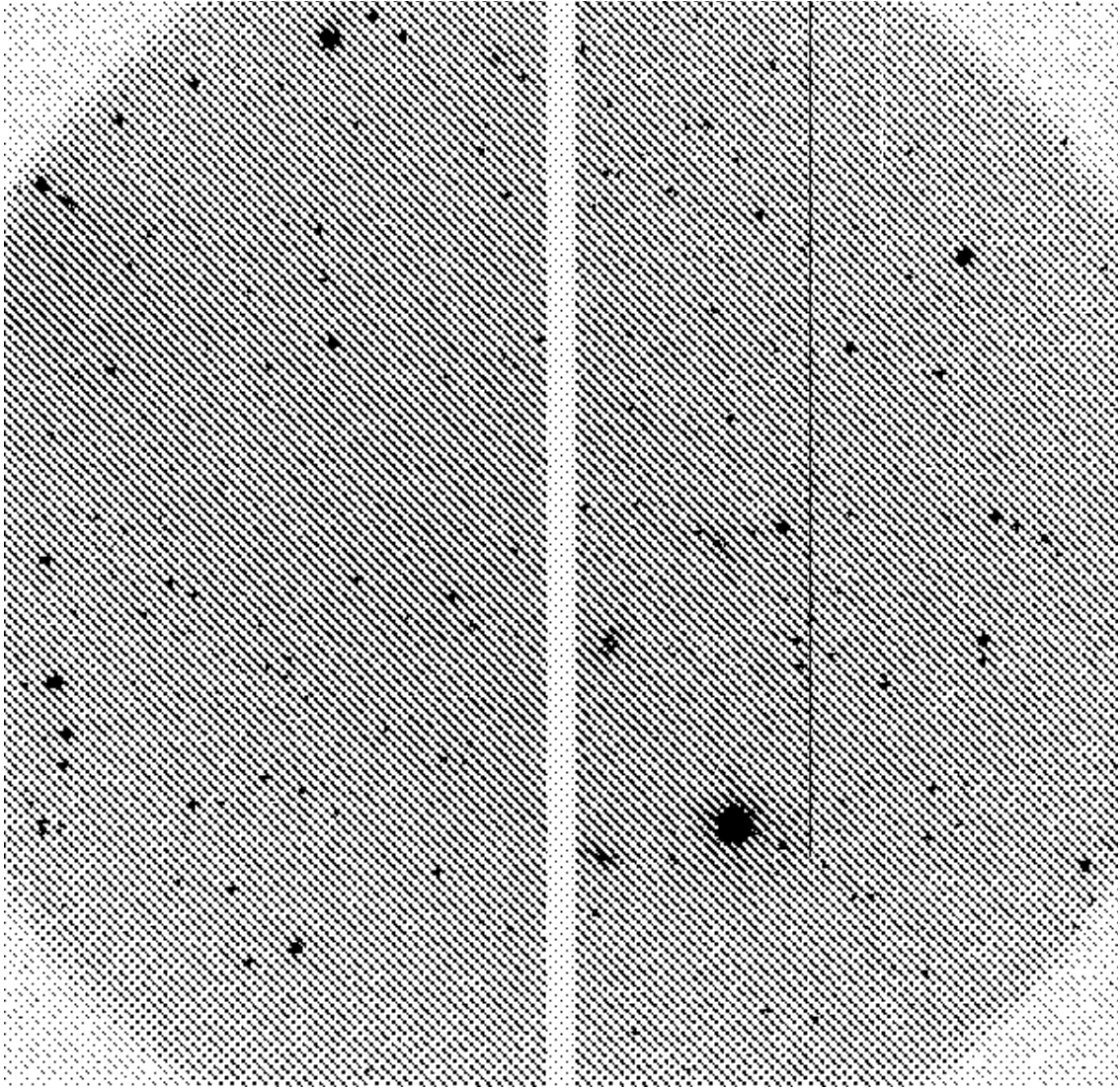}
}
 \caption{V-band image of \AM\ with SALTICAM. This is a 2 sec
 exposure, primarily reduced, using binning on-chip of 2$\times$2
 pixels (to 0.28 arcsec) and without fixing cosmetic defects such as a few
 bad columns. The full image is displayed here to emphasize the
 full extent of the imaged area.
 \AM\ is just below and to the right of the centre of the image. PGC 400092 is to
 its upper-right (North-West) and PGC 399718 is below (South), next to the
 bright star. The vertical size of the image is 575 arcsec. The SALTICAM
 science field covers 480 arcsec and the outer 60 arcsec annulus is,
 in principle, used for guiding.
 \label{fig:AM_direct}}
\end{figure*}

\begin{figure*}
{\centering
 \includegraphics[clip=,angle=0,width=12.0cm]{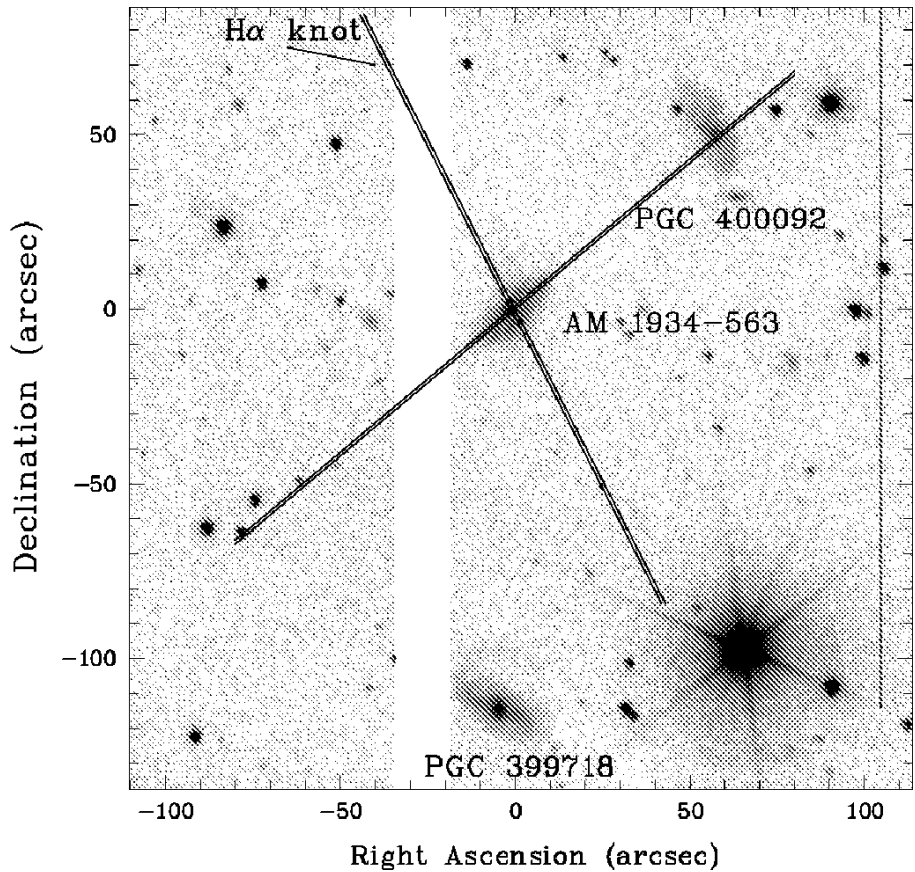}
}
 \caption{A 220$\times$220 arcsec image extracted from the one shown in Fig.~\ref{fig:AM_direct}.
 The three galaxies of the tight group are indicated, as is the newly
 detected H$\alpha$ knot (see text for more details).
 The slit positions used here are over-plotted and each slit is 1.5\arcsec
 wide.Note a few other diffuse images in the neighbourhood.
 \label{fig:AM_direct_slits}}
\end{figure*}

\begin{figure*}
{\centering
 \includegraphics[clip=,angle=-90,width=17.0cm]{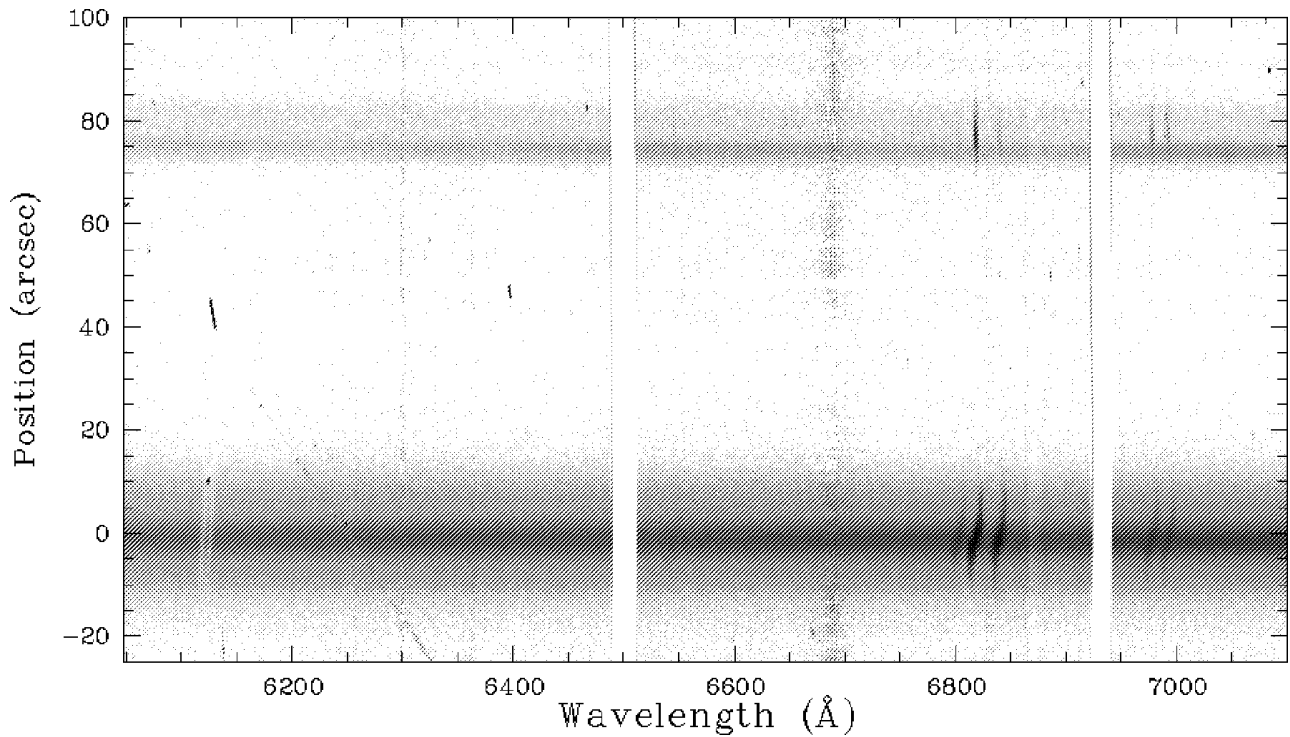}
}
 \caption{%
Part of 2D reduced spectrum  for $\rm PA = 140\degr$. NW is up.
The slit was positioned along the major axis of \AM\ and exhibits
the redshifted H$\alpha$, [\ionn{N}{ii}] $\lambda\lambda$6548,
6583, and [\ionn{S}{ii}] $\lambda\lambda$6716,6731 emission lines
with measurable intensities. The \Na\,D $\lambda\lambda$5890, 5896
absorption doublet can easily be seen. The spectrum of \AM\ is
visible for a distance of $\pm$15\arcsec\ along the slit, but the
emission and absorption lines can be reliably traced only up to
$\pm$10\arcsec. The spectrum of the Sd/Irr galaxy PGC~400092 is
located approximately $\sim$80\arcsec\ away from \AM\,. The
PGC~400092 spectrum shows the same emission lines as \AM, but
there is no indication of \Na\ absorption. Weak [\ionn{O}{i}]
$\lambda$6300 and \ionn{He}{i} $\lambda$5876 lines are also
present. At the adopted distance of 167 Mpc, 1\arcsec\ = 0.8 kpc
and the image extent is $\sim$100 kpc.
    \label{fig:AM_2D_130}}
\end{figure*}

\begin{figure*}
{\centering
 \includegraphics[clip=,angle=-90,width=17.0cm]{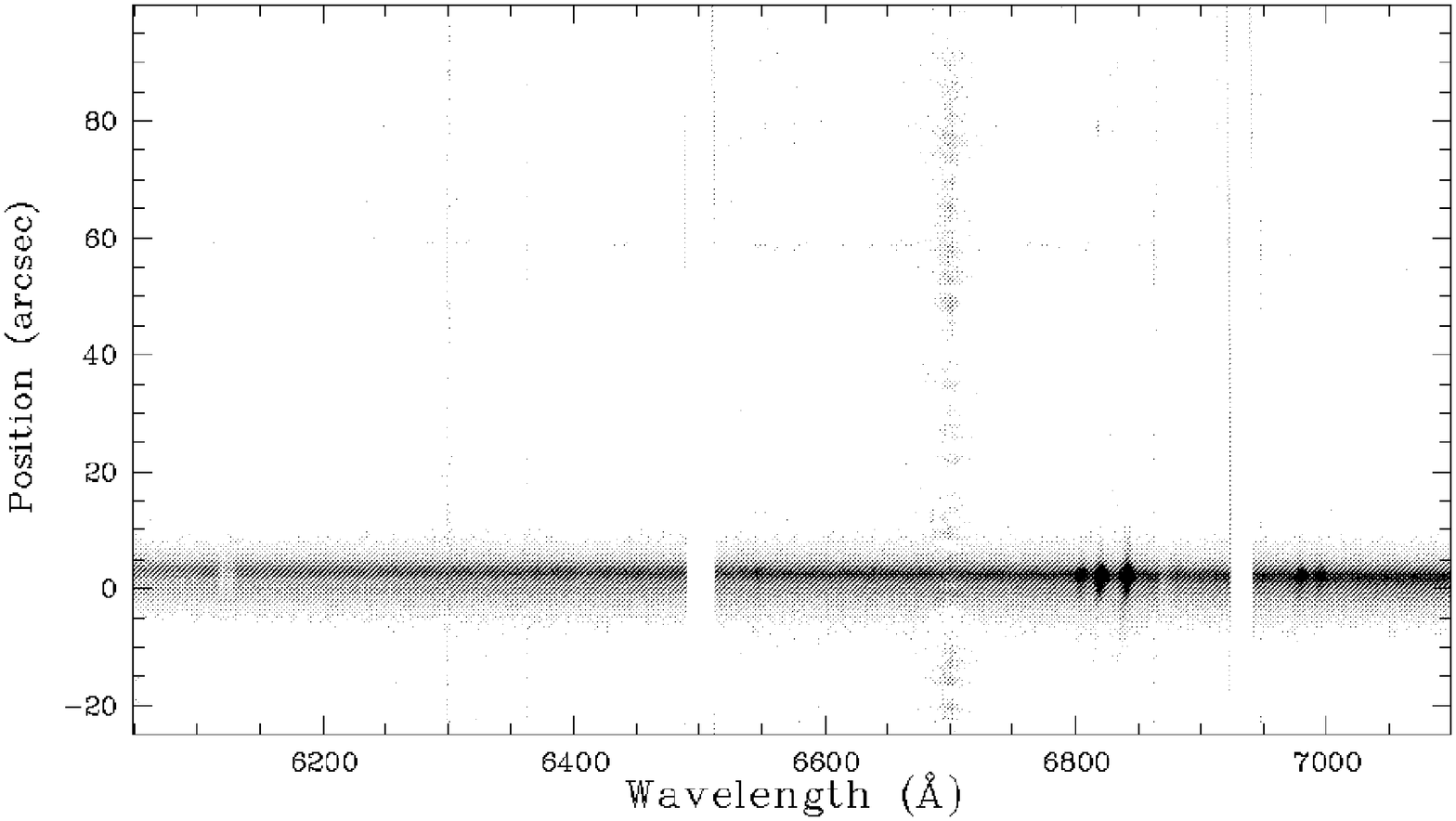}
}
 \caption{%
Part of  2D reduced spectrum obtained at $\rm PA= 27\degr$\, that covers the
same spectral region as the spectrum for $\rm PA= 140\degr$\, and
shows the same spectral features. NE direction is up. The slit is
positioned along the major axis of the polar ring of \AM. The
spectrum of \AM\, is visible at position 0$\pm$10\arcsec. Note the
H$\alpha$ emission line produced by the newly detected group
member $\sim$80\arcsec\, away from  \AM\,, near the top and close
to the right edge of the image. The linear scale and extent of
this images are identical to those of Fig.~\ref{fig:AM_2D_130}.
    \label{fig:AM_2D_27}}
\end{figure*}

\section{Observations and data reduction}
\label{txt:Obs_and_Red}

SALT was described by
Buckley \etal (2006) and by O'Donoghue \etal (2006), its Robert
Stobie Spectrograph (RSS) was described by Burgh \etal (2003) and Kobulnicky et al.
(2003), and
the first scientific papers based on its observations were
published by Woudt \etal (2006) and by O'Donoghue \etal (2006). We
used the SALT and RSS to observe \AM. The observations of \AM\
were obtained during the Performance Verification (PV) phase of
the SALT telescope with the RSS spectrograph and are described in
Table \ref{t:Obs}.

The July 2006 spectra (see Table~\ref{t:Obs}) were obtained during
unstable weather conditions (high humidity, seeing worse than
5\arcsec), without fully stacking the SALT mirrors. They cover the
range 3650--6740 \AA\ with a spectral resolution of $\sim$1.0
\AA\, pixel$^{-1}$ or a FWHM of 6--7 \AA. These spectra do not
show strong and extended emission lines but were used to
measure equivalent widths (EWs) of absorption lines in that
spectral range following observations.

The spectra obtained on the nights of September 2006 were taken
during stable weather conditions with seeing $\sim$1\farcs5. They
cover the range from $\sim$6050\AA\, to $\sim$7300\AA\,  with a
spectral resolution of 0.4 \AA\, pixel$^{-1}$ or 2.4 \AA\ FWHM.
All data were taken with a 1\farcs5 wide slit and a final
scale along the slit of 0\farcs258 pixel$^{-1}$ (after binning the
CCDs by a factor of two). Each exposure was broken up into 2--3
sub-exposures to allow the removal of cosmic rays. Spectra of a
Cu--Ar comparison lamp were obtained after the science exposures
to calibrate the wavelength scale.

The September 2006 data include two spectra obtained at position
angle 140$^{\circ}$ centered on \AM\, extending about four arcmin
along the galaxy's major axis and at a shallow angle to the dust
lane, where the northern part passes also through the ``northwest
companion'' PGC 400092 \citep{Resh06}, and three spectra centered
on the same position but obtained at position angle 27$^{\circ}$,
along the major axis of the ''polar ring'' described by
\citet{Resh06}. We emphasize that the sampling of the major axis
spectra was at PA=140$^{\circ}$, not at 130$^{\circ}$ as done by
\citet{Resh06}, since 140$^{\circ}$ is closer to the position
angle of the disk as given by Reshetnikov \etal (148$^{\circ}$)
and allows for a moderate degree of disk warping.

Although the observations discussed here are mostly spectroscopic,
one image of the galaxy was obtained with a two-sec exposure in
the V filter with the SALTICAM camera (O'Donoghue \etal 2006)
prior to the spectrometer observations in order to adjust the slit
orientation, and is shown here as Figure~\ref{fig:AM_direct}. The
$\sim$1\farcs5 seeing during the observations, and the problematic
image quality SALT exhibited at that time, which can be evaluated
from the stellar images on Figure~\ref{fig:AM_direct_masked} (see
below), caused the images far from the good-quality $\sim$3 arcmin
region to assume complicated shapes. The full SALTICAM image is
$\sim$10 arcmin across with 0.28 arcsec/pixel (after binning
on-chip by a factor of two).

The data for each RSS chip were bias and overscan subtracted, gain
corrected, trimmed and cross-talk corrected, sky-subtracted and
mosaiced. All the primary reduction was done using the
IRAF\footnote{IRAF: the Image Reduction and Analysis Facility is
distributed by the National Optical Astronomy Observatory, which
is operated by the Association of Universities for Research in
Astronomy, In. (AURA) under cooperative agreement with the
National Science
Foundation (NSF).} package {\it salt}\footnote{%
See
http://www.salt.ac.za/partners-login/partners/data-analysis-software
for more information.} developed for the primary reduction of SALT
data. Cosmic ray removal was done with the FILTER/COSMIC task in
MIDAS.\footnote{MIDAS is an acronym for the European Southern
Observatory package -- Munich Image Data Analysis System.} We used
the IRAF software tasks in the {\it twodspec} package to perform
the wavelength calibration and to correct each frame for
distortion and tilt. One-dimensional (1D) spectra were then
extracted using the IRAF APALL task.

\begin{figure*}
{\centering
 \includegraphics[clip=,angle=-90,width=15.5cm]{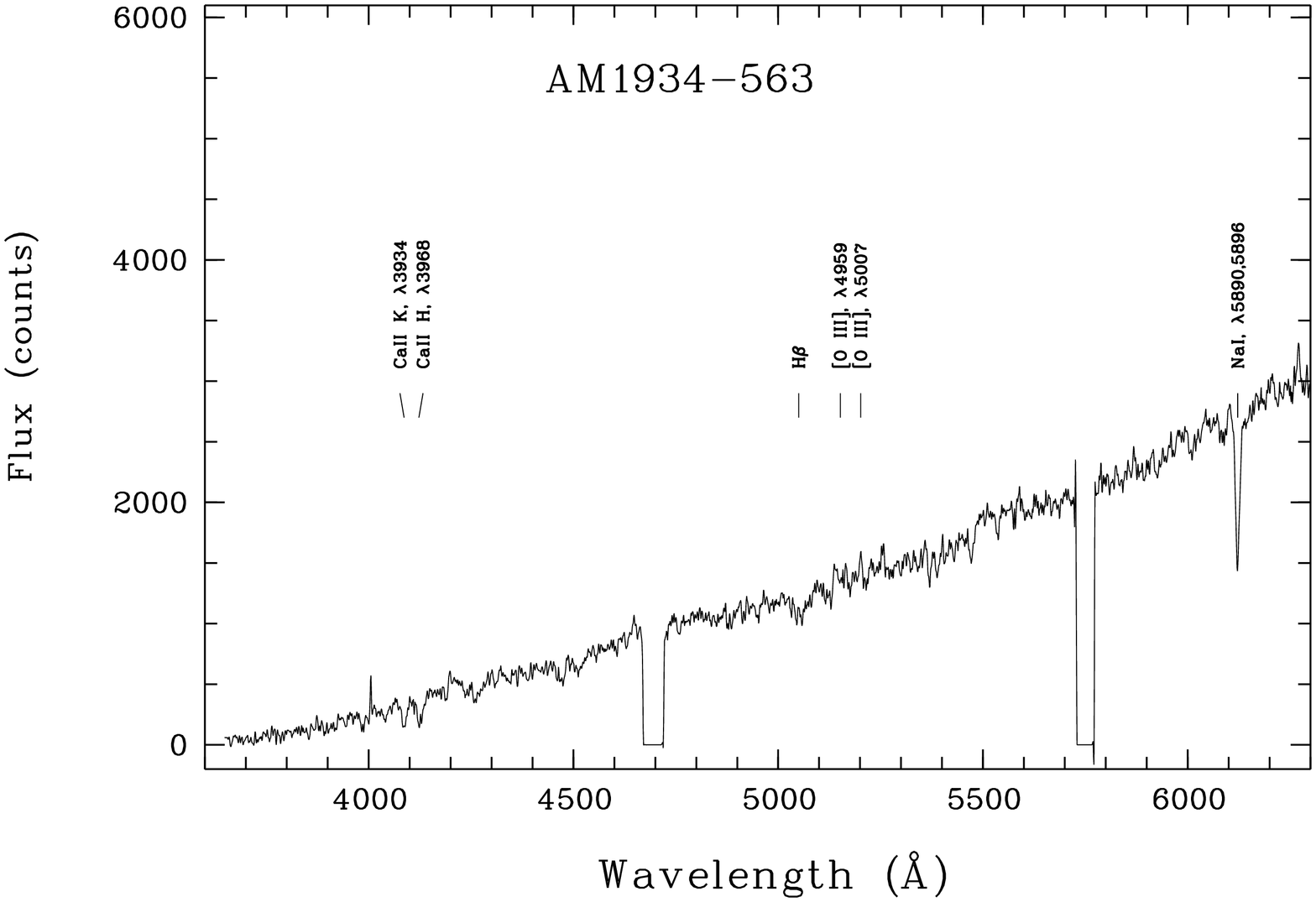}
 \includegraphics[clip=,angle=-90,width=15.5cm]{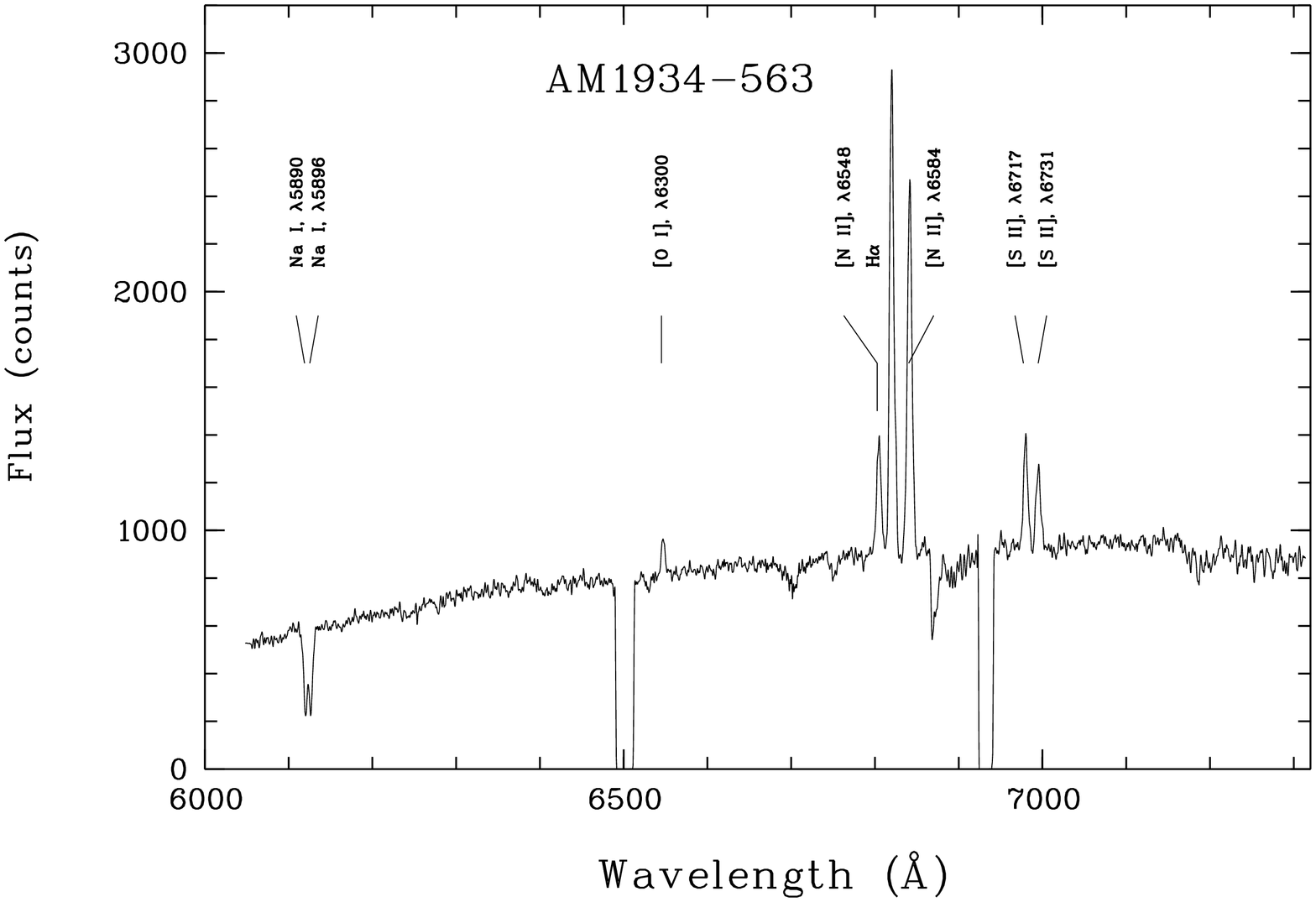}
}
 \caption{%
{\it Top panel:} The 1D spectrum of the central part of \AM\,
extracted from the 2D spectrum observed at $\rm PA= 140\degr$\
with a setup that covers 3650--6740\AA\, and with a spectral scale
of $\sim$1\AA\, pix$^{-1}$. The ``reddest'' part of the spectrum
is not shown. The spectrum shows some absorption lines and
possibly very weak [\ionn{O}{iii}] $\lambda\lambda$4959, 5007
emission lines. {\it Bottom panel:} The 1D spectrum of the central
part of \AM\, extracted from the 2D spectrum observed at $\rm PA=
27\degr$. All detected lines have been marked.
    \label{fig:AM_1D_130}}
\end{figure*}

\begin{figure*}
{\centering
 \includegraphics[clip=,angle=-90,width=17.0cm]{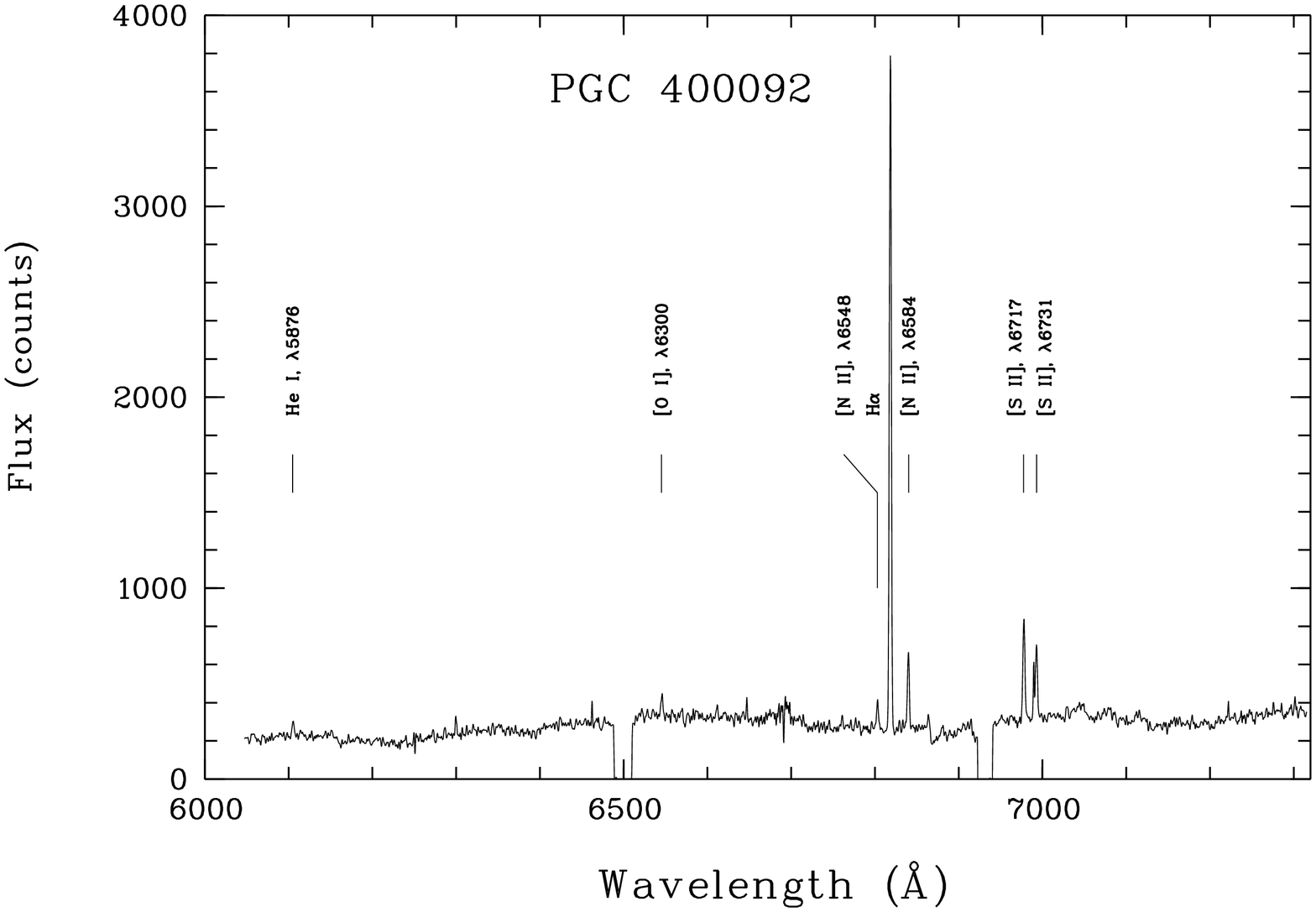}
}
 \caption{%
The 1D spectrum of PGC~400092, extracted from the 2D spectrum
observed at $\rm PA = 140\degr$. All the detected emission lines have been
marked. Note that no \Na\,D $\lambda\lambda$5890, 5896 absorption
lines are visible in this spectrum and the continuum is very weak
in comparison with the \AM\, spectrum shown in the bottom panel of
Fig.~\ref{fig:AM_1D_130}.
    \label{fig:PGC40092_1D}}
\end{figure*}

Figures~\ref{fig:AM_2D_130} and \ref{fig:AM_2D_27} show parts of
fully reduced and combined spectral images for PA=140$^{\circ}$
and PA=27$^{\circ}$, respectively. Figure~\ref{fig:AM_1D_130}
shows the spectrum of the central part of \AM. The $\sim$40\AA\,
missing sections at $\sim \lambda \lambda$ 6500 and 6930\AA\, are
produced by small gaps between the three CCDs of the RSS. The
noisy region of the RSS images shown in Figs.~\ref{fig:AM_2D_130}
and \ref{fig:AM_2D_27} near $\sim$6685\AA\, is a subtraction
artifact of laser light scattered into the RSS from SALT's
interferometric auto-collimating system.
Figure~\ref{fig:AM_1D_130} shows the 1D spectra of the central
part of \AM\, extracted from the 2D spectra.
Figure~\ref{fig:PGC40092_1D} shows the 1D spectrum of the galaxy
PGC~400092 extracted from the 2D spectrum observed at PA=140\degr.

The derived internal errors for the 2D wavelength calibrations
were small and did not exceed 0.04 \AA\ for a resolution of 0.4
\AA\ pixel$^{-1}$, or $<$2 km s$^{-1}$ at the wavelength of
redshifted H$\alpha$ line. To exclude systematic shifts
originating from known RSS flexure, we calculated line-of-sight
velocity distributions along the slit for both emission and
absorbtion lines using a suite of MIDAS programs described in
detail in \citet{Zasov00}. These programs allow the use of
additional correction factors derived from tracing nearby
night-sky lines whose accurate wavelengths are very well known to
correct the observed wavelengths of the \Na\,D, H$\alpha$\,
[\ionn{N}{ii}] $\lambda$6583 and [\ionn{S}{ii}] $\lambda$6716
emission lines. After implementing the night-sky line corrections,
the line-of-sight velocity distributions are accurate to $\sim$1.5
km s$^{-1}$. Most of the calculated velocity distributions are
shown in Figures~\ref{fig:AM_rot_130a}--\ref{fig:AM_rot_27}.
All velocities derived with this procedure are heliocentric.

All emission lines were measured with the MIDAS programs described
in detail in \citet{SHOC, Sextans}. These programs determine the
location of the continuum, perform a robust noise estimation, and
fit separate lines with single Gaussian components superposed on
the continuum-subtracted spectrum. Nearby lines, such as the
H$\alpha$  and [\ionn{N}{ii}] $\lambda\lambda$6548, 6583 lines on
the one hand, the [\ionn{S}{ii}] $\lambda\lambda$6716, 6731 lines
on the other, and \Na\,D $\lambda\lambda$5890, 5896 absorption
doublet were fitted simultaneously as blends of two or more
Gaussian features.

\section{Results}
\label{txt:results}

\subsection{Spectra of \AM\ and PGC~400092}

A cursory inspection of the spectra obtained at PA=140$^{\circ}$
(see Figure~\ref{fig:AM_2D_130}) shows rotation detectable in the
same amount and behaviour exhibited by the H$\alpha$,
[\ionn{N}{ii}] $\lambda\lambda$6548,6583 and [\ionn{S}{ii}]
$\lambda\lambda$6716,6731 emission lines, and rotation as almost
a solid body exhibited by the \Na\ $\lambda\lambda$5890,5896
doublet lines. The NE extension of the spectrum, away from \AM\
and crossing the companion galaxy PGC~400092, shows that the same
emission lines seen in \AM\ are produced by the NE companion; the
rotation there is much slower and the \Na\ doublet is not visible,
even though the continuum there is visible. In addition, the
spectrum of PGC~400092 shows also weak [\ionn{O}{i}] $\lambda$6300
and HeI $\lambda$5876 in emission, while the spectrum of \AM\,
shows [\ionn{O}{i}] $\lambda$6300 emission only in the central
part.

\begin{table}
\caption{EWs of absorption lines in spectra of \AM}
\label{t:EW_abs}
\begin{center}
\begin{tabular}{lc} \hline
  Absorption Line   &  Equivalent Width  \\
  \MC{1}{c}{(\AA)}  &  (\AA)             \\ \hline
  CaII H            & ~8.9$\pm$1.5       \\
  CaII K            & 10.3$\pm$1.8       \\
  H$\delta$         & ~6.5$\pm$2.1       \\
  H$\gamma$         & ~5.8$\pm$2.4       \\
  H$\beta$          & ~6.4$\pm$2.5       \\
  \Mg\,b            & ~3.5$\pm$0.8       \\
  \Na\,D            & ~5.8$\pm$0.7       \\  \hline
\end{tabular}
\end{center}
\end{table}

The short-wavelength spectra obtained in June 2006 (top panel of
Fig.~\ref{fig:AM_1D_130}) show the blend of the \Na\ doublet as a
single line (due to the lower resolution of this setup), and the
H$\beta$, H$\gamma$ and H$\delta$ lines in absorption. The CaII H
and K doublet is seen in absorption at the blue end of the
spectrum. The spectra also show very weak [\ionn{O}{iii}]
$\lambda\lambda$4959, 5007 emission lines. In this figure and in
the following plots we describe as ''intensity'' the raw counts
extracted from the spectra. Since our data have not been
spectrophotometrically calibrated, this is in reality ''relative
intensity''. The equivalent widths of the main  absorption
lines were measured for the central part of the galaxy and are
shown in Table~\ref{t:EW_abs}. Measurements of lines detected in
more than one spectrum were averaged.

\begin{figure*}
{\centering
 \includegraphics[clip=,angle=-0,width=15.0cm]{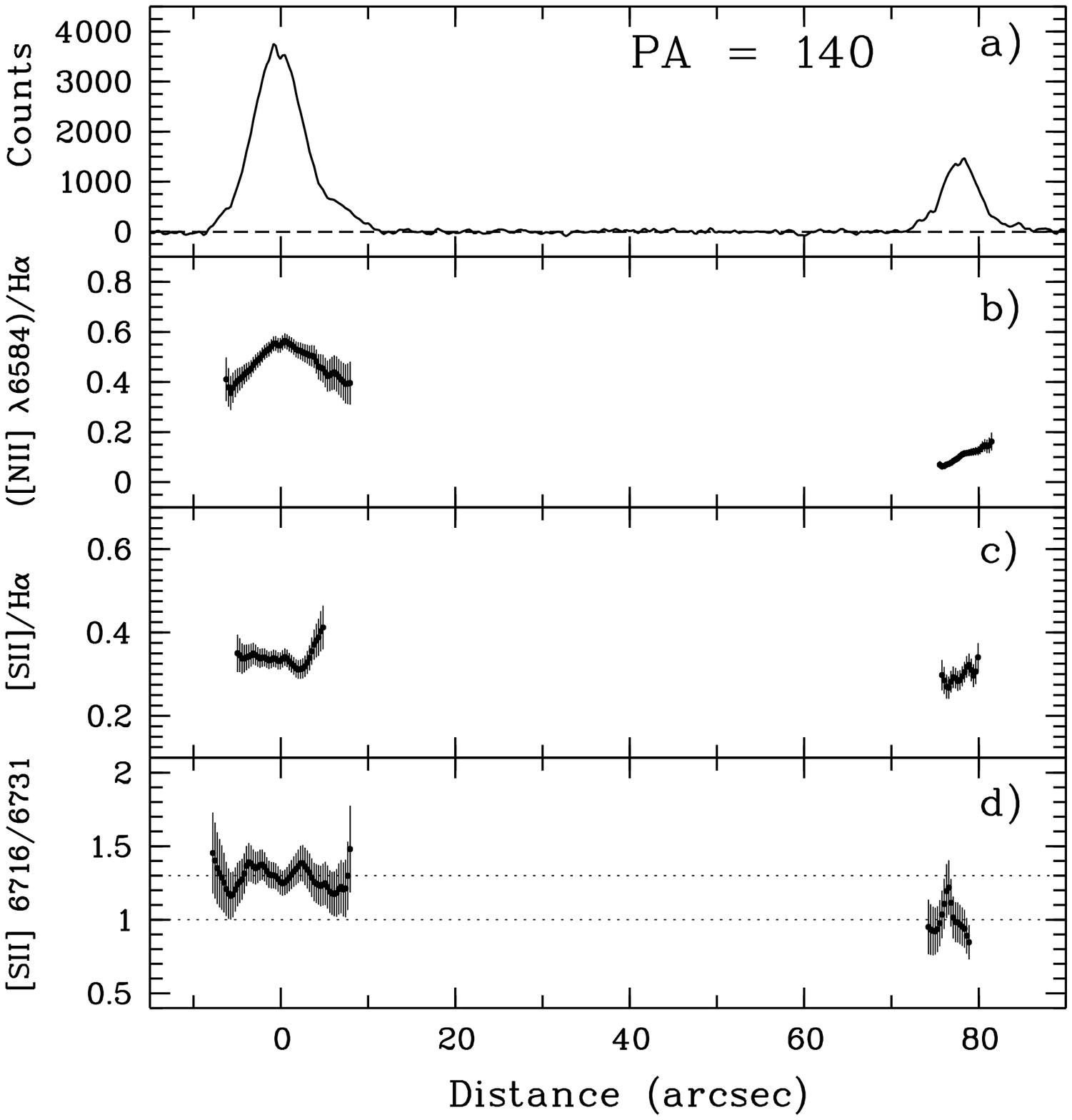}
}
 \caption{%
Line count ratios along the slit for $\rm PA = 140\degr$.
All points displayed here have a signal-to-noise ratio of at least
four. {\it Top to bottom:} a). Profile of the net H$\alpha$ flux
in total counts. b). Profile of the [\ionn{N}{ii}]
$\lambda$6583/H$\alpha$ ratio. c). Profile of the [\ionn{S}{ii}]
6716+6731/H$\alpha$ ratio. d). Profile of the electron-density
sensitive ratio R$_{SII}$=[\ionn{S}{ii}]6716/[\ionn{S}{ii}]6731.
The value R$_{SII}$=1.4 is plotted with a dotted line. The
values R$_{SII}$=1.35 and 1.0 are plotted with dotted lines; these
indicate electron densities $\rm n_e=50$ and 500 cm$^{-3}$
respectively.
    \label{fig:AM_140_cond}}
\end{figure*}

\begin{figure*}
{\centering
 \includegraphics[clip=,angle=0,width=13.0cm,height=16cm]{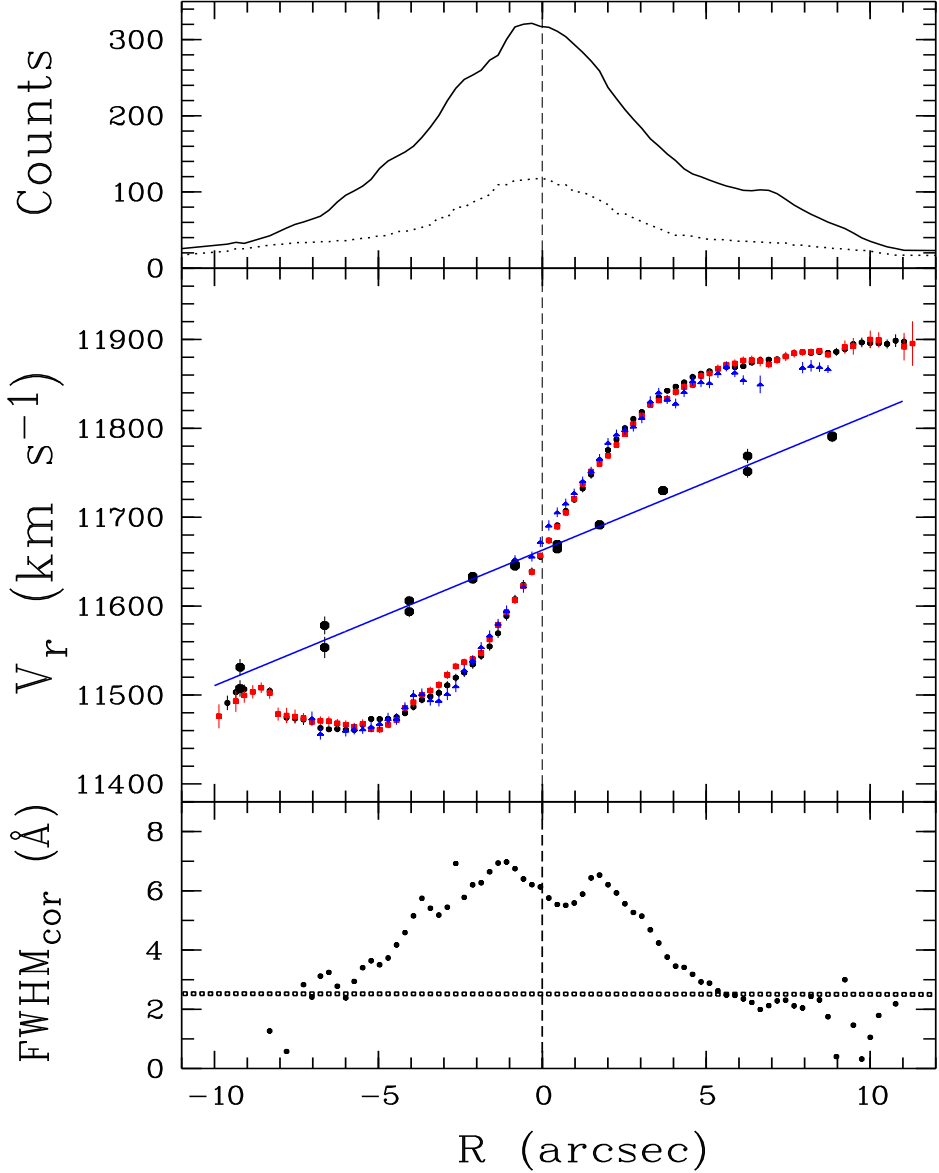}
}
 \caption{%
{\it Top panel:} The solid line shows the profile of the H$\alpha$
flux along the slit for $\rm PA = 140\degr$\, after continuum
subtraction. The short-dashed line shows the continuum intensity
distribution along the slit and in the spectral region of the
H$\alpha$ line. {\it Middle panel:} Radial velocity distribution
along the major axis of \AM. The black squares, red squares and
blue triangles represent measurements of the emission lines
H$\alpha$, [\ionn{N}{ii}] $\lambda$6583 and [\ionn{S}{ii}]
$\lambda$6716 respectively. The black filled circles show the
stellar velocity distribution measured from the absorption doublet
\Na\,D $\lambda\lambda$5890, 5896. One $\sigma$ error bars have
been overplotted for all measurements. The solid blue line is
result of a linear fit to all measurements of the \Na\,D lines.
{\it Bottom panel:}
The measured FWHM of the H$\alpha$ line, corrected for the
intrinsic line width of the RSS. The FWHM of the reference
night-sky line measured in each row is shown with open squares.
    \label{fig:AM_rot_130a}}
\end{figure*}

\begin{figure*}
{\centering
 \includegraphics[clip=,angle=-90,width=16cm]{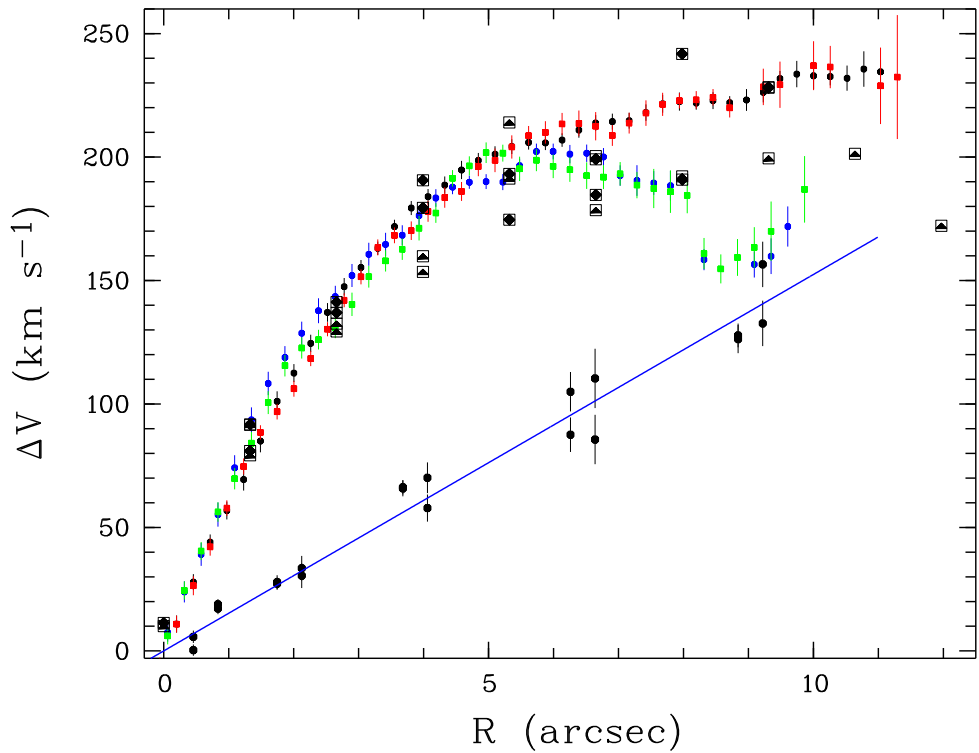}
}
 \caption{%
The galacto-centric velocity distributions along the major axis of
\AM. The black and red filled circles are for the NW branch using
the emission lines of H$\alpha$ and [\ionn{N}{ii}] $\lambda$6583
lines, respectively. The blue and green filled circles are for the
SE branch using the H$\alpha$ and [\ionn{N}{ii}] $\lambda$6583
lines. The filled black circles show the stellar velocity
distribution measured from the absorption doublet \Na\,D
$\lambda\lambda$5890, 5896. The solid blue line is the result of a
linear fit to all measurements for the \Na\,D lines (see
Section~\ref{txt:results} for details).
Big black filled lozenges and triangles placed into squares represent the \citet{Resh06} data
for H$\alpha$ and [\ionn{N}{ii}] $\lambda$6583, respectively.
These values have not been corrected back for cosmological
stretch.
    \label{fig:AM_rot_130b}}
\end{figure*}

\begin{figure*}
{\centering
 \includegraphics[clip=,angle=-0,width=15.0cm]{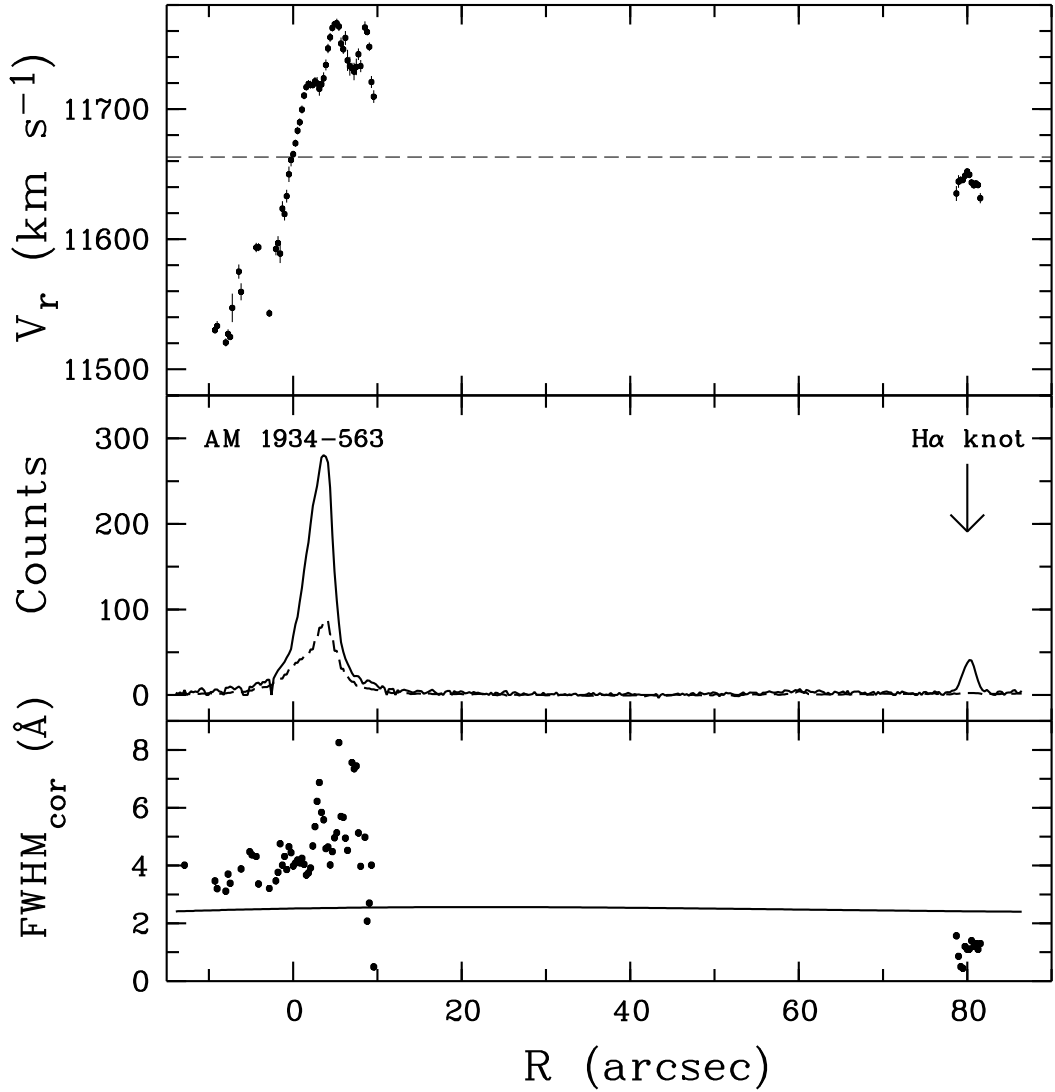}
}
 \caption{%
{\it Top panel:} The radial velocity distribution of the H$\alpha$
emission line along the major axis of the ring of \AM. The
H$\alpha$ emission line  produced by the newly detected group
member appears $\sim$80\arcsec\ away from the center of \AM\,.
This newly detected group member has a small velocity dispersion
and only a $\sim$20 km s$^{-1}$ difference from the systemic
velocity of \AM\,, which is plotted with a short-dashed line. {\it
Middle panel:} The solid line shows the profile of the H$\alpha$
flux along the slit at $\rm PA = 27\degr$\, with the continuum
subtracted. The short-dashed line shows the continuum intensity
distribution in the region of the line and along the slit. {\it
Bottom panel:} The measured FWHM for H$\alpha$ line corrected for
the RSS intrinsic line width. The FWHM of the reference night-sky
line is shown as the solid line.
    \label{fig:AM_rot_27_all}}
\end{figure*}

\begin{figure}
{\centering
 \includegraphics[clip=,angle=-0,width=8.5cm]{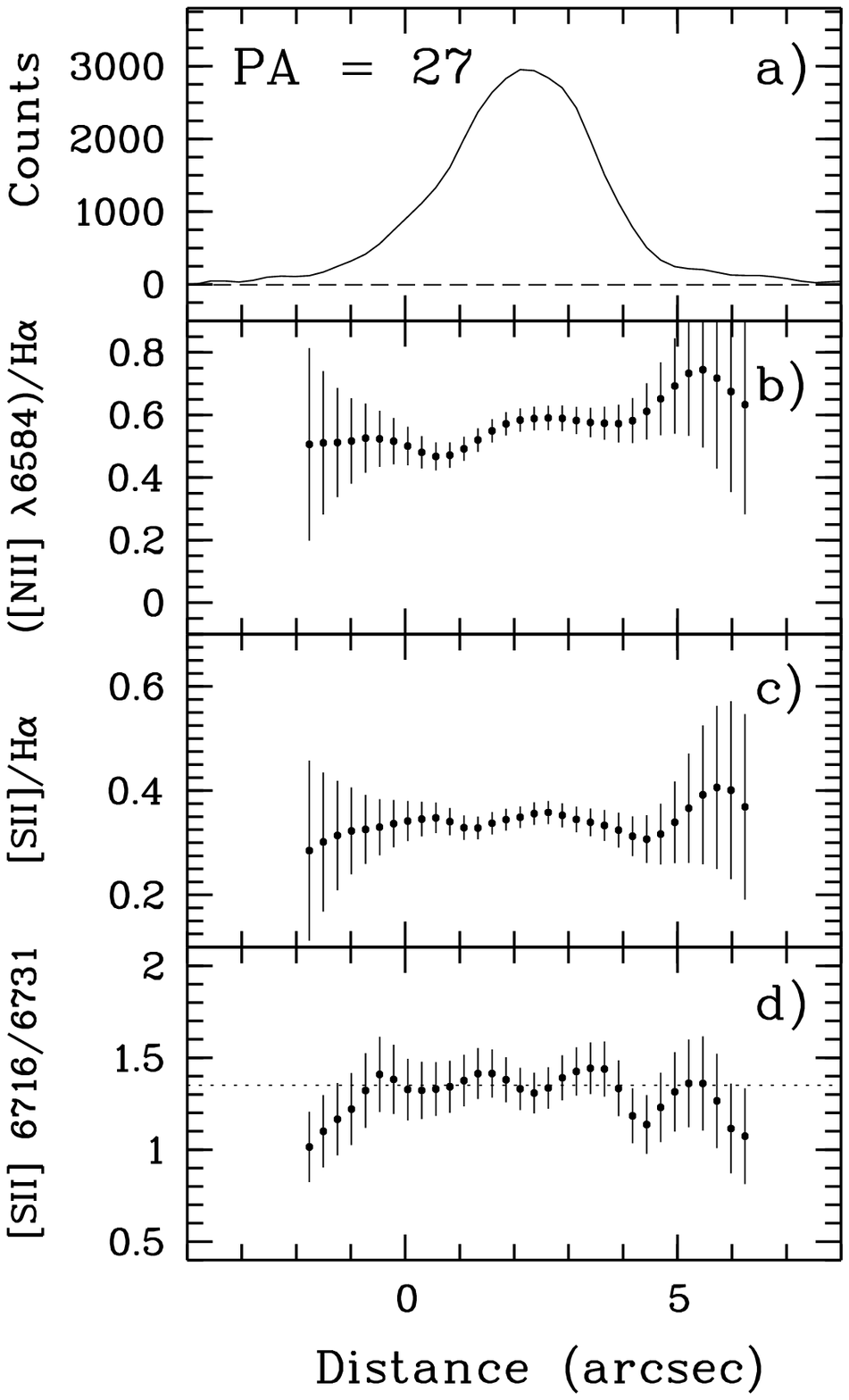}
}
 \caption{%
Line count ratios along the slit for $\rm PA = 27\degr$. All
points plotted here have a signal-to-noise ratio of at least four.
{\it Top to bottom:} a). Profile of the H$\alpha$ flux in total
counts. b). Profile of the [\ionn{N}{ii}] $\lambda$6583/H$\alpha$
ratio.  c). Profile of the [\ionn{S}{ii}] 6716+6731/H$\alpha$
ratio. d). Profile of the electron-density sensitive
[\ionn{S}{ii}]6716/[\ionn{S}{ii}]6731 R$_{[\ionn{S}{ii}]}$ ratio.
The value R$_{[\ionn{S}{ii}]}$=1.35 is plotted with a dotted line
and indicates an electron density $\rm n_e=50$ cm$^{-3}$.
    \label{fig:AM_27_cond}}
\end{figure}

The rotation curve of \AM\ along the major axis, derived from the
two-spectra combination shown in Figure~\ref{fig:AM_2D_130}, is
shown in Figures~\ref{fig:AM_rot_130a}
and \ref{fig:AM_rot_130b}. Figure~\ref{fig:AM_rot_130a} show
the velocity-position plot and
Figure~\ref{fig:AM_rot_130b} shows the galacto-centric
velocity-distance plot. In general, the emission-line rotation
curve derived here corresponds with that shown in Figure~5 of
\citet{Resh06}, except that ours is better sampled, has a higher
signal-to-noise, and the rotation curves derived from the
different emission lines practically coincide, as can be estimated
from the formal 1$\sigma$ error bars plotted in the figures and
from the scatter of the individual points.
Figure~\ref{fig:AM_rot_130b} shows also a comparison of our
measurements with those of \citet{Resh06}.

Deriving the rotation curves shown in
Figures~\ref{fig:AM_rot_130a} and \ref{fig:AM_rot_130b} we found
that the systemic radial velocity of \AM\ is 11663$\pm$3 km
sec$^{-1}$, formally higher by some 14 km sec$^{-1}$ than the
value given by \citet{Resh06} in their Table~3 but consistent with
their value within the quoted uncertainties. This offset might be
the result of a slightly different definition of the systemic
velocity; we chose the value for which the NW branch of the
rotation curve matched best that for the SE branch and by this
procedure also found the rotation center of the galaxy.
Independently, we found that this location on the velocity curve
is also the central point for the linear fitting of all the
measurements for the \Na\,D lines seen in absorption, as shown in
Figure~\ref{fig:AM_rot_130a}. We obtained a best-fit line
following the relation:
\begin{equation}
{\rm V_r = (11663 \pm 2) + (15.2 \pm 0.4) \times {\rm R} } 
\end{equation}
where $\rm R$ is the distance in arcsec from the point where the
radial velocity of \AM, defined using the emission lines, equals
11663 km sec$^{-1}$ and we adopt this location as the kinematic
centre of the galaxy.
The different symbols indicate the H$\alpha$ velocity (black
squares), the [\ionn{N}{ii}] $\lambda$6583 line velocity (red
squares), and the [\ionn{S}{ii}] $\lambda$6716 line velocity (blue
triangles). The stellar rotation along the same slit position on
the major axis, as derived from an average of the two \Na\
absorption lines, is depicted as filled black circles.

We detected a discrepant systemic velocity 11680$\pm$10 km sec$^{-1}$
for the NW companion PGC~400092 as well,
where our value is significantly lower than the
11735$\pm$6 km sec$^{-1}$ given in \citet{Resh06}. Since the
velocity discrepancies for \AM\ and for PGC\,400092 are in opposite
directions, we can probably rule out a systematic shift between
our velocity scale and the one of \citet{Resh06}. This is confirmed also
by the plot in Figure~\ref{fig:AM_rot_130b} where their derived velocity
curve points are plotted over our results. The shift
between our data for PGC\,400092 and that from \citet{Resh06} could
be the result of the slit position for $\rm PA = 140\degr$\ used
here that did not cross exactly the physical center of that galaxy.

We could also derive the velocity dispersion of the H$\alpha$ line
along the slit for $\rm PA = 140\degr$; this is shown in the
bottom panel of Figure~\ref{fig:AM_rot_130b}. The dispersion is
shown as the FWHM of the line after correcting for the intrinsic
spectrometer line width. The corrected H$\alpha$ line
FWHM=5--7 \AA\ found for the central part ($\pm$3 arcsec) of \AM\
indicates internal motions of 200--300 km s$^{-1}$. The corrected
FWHM=$<$1 \AA\ measured for the H$\alpha$ line of PGC\,400092
indicates internal motions slower than 45 km s$^{-1}$.

The rotation curve along the polar ring axis, at $\rm PA =
27\degr$, is shown in Figure~\ref{fig:AM_rot_27} as a
velocity-position plot. This, as already mentioned, relies mostly
on the emission lines since the \Na\ absorptions are visible only
in the central part of the spectrum, and is therefore more limited
in extent. The spectra for $\rm PA = 27\degr$\, show a linearly
increasing rotation for $\sim$7\arcsec\ SW of the galaxy centre
outwards, where the center position is that derived for the major
axis. Since the NE and SW branches of the ring's major axis show
very different behaviour from that observed along the galaxy's
major axis, the method used previously to find the rotation center
by matching the two branches could not be used in this case, thus
we do not show a folded and combined velocity curve for the major
axis of the ring.

The NE branch shows an approximately flat rotation from
$\sim$2\arcsec\ away from the centre, as derived from the emission
lines, with some oscillations from the center to the periphery at
10 arcsec from the center. These oscillations are evident in both
H$\alpha$ and [\ionn{N}{ii}] $\lambda$6583; they may be caused by
the overlap of the emission lines from the ring with those from
the main body of the galaxy. The plot in the top panel of
Fig.~\ref{fig:AM_rot_27} shows that the strongest H$\alpha$
emission is encountered close to the location of the most intense
continuum contribution (compare the solid and the dashed lines).

Our spectra along $\rm PA = 27\degr$ show a completely different
kinematic behaviour than the one described by \citet{Resh06}.
Their Fig.~7 shows a $\sim$50 km sec$^{-1}$ difference between the
velocity of the [\ionn{N}{ii}] $\lambda$6583 and H$\alpha$ at the
galaxy centre that increases to $\sim$100 km sec$^{-1}$ at the SW
end of the ring. We, on the other hand, see no difference between
the velocities of these two lines. Moreover, the [\ionn{S}{ii}]
lines in our observed spectrum also show the same behavior as the
[\ionn{N}{ii}] $\lambda$6583 and H$\alpha$ lines. We also note
that the extent to which the rotation is defined and measurable
for this position angle and using the emission lines is
practically the same as for the major axis of \AM\,, some 8 kpc
from the center (at 167 Mpc).

Similar to the case of the major axis, $\rm PA = 140\degr$, we see
here also a straight-line behaviour with galacto-centric distance
of the \Na\ absorption lines. We find a formal linear fit of the
form
\begin{equation}
{\rm V_r = (11662 \pm 2) + (14.9 \pm 0.8) \times {\rm R}} 
\end{equation}
The \Na\ rotation curve is linear from 1\farcs5 SW of the centre to
$\sim$5" NE of the kinematic centre. Note that the value found for
the slope at this position angle is virtually identical with that
for the major axis in equation (1).

A comparison of the two panels of Fig.~\ref{fig:AM_rot_27}, the
lower one which is a velocity-position plot for $\rm PA = 27\degr$
and the upper one which is a plot of the line intensity vs.
position along the slit, shows that the region where most of the
line emission is produced is about 4\arcsec\ to the NE of the
kinematic center of \AM\ and that the emission is practically only
along the NE part of the ring.

As for $\rm PA = 140\degr$, we derive the velocity dispersion for
this position angle as the FWHM of the H$\alpha$ line vs.
galacto-centric distance. This is shown in the bottom panel of
Fig.~\ref{fig:AM_rot_27_all} after correction for the intrinsic
width of the lines using the night sky spectrum. The corrected
FWHM=7 \AA\ for the redshifted H$\alpha$ indicates internal
motions of $\sim$300 km s$^{-1}$.

Although not spectrophotometrically calibrated, our spectra allow
the derivation of a few physical parameters of the gas using line
ratios. The good signal-to-noise of the spectra allows the
derivation of these ratios along the slit, as shown in
Figs.~\ref{fig:AM_140_cond}  and~\ref{fig:AM_27_cond}. The ratios
plotted in Fig. \ref{fig:AM_140_cond} allow a derivation along the
galaxy major axis and for its NW companion. Since these
ratios are based on the very closely located emission lines, they
practically do not depend on whether the spectral data were
corrected for sensitivity or not. For the red spectral range,
using the sensitivity curve cannot change these ratios by more
that a few percent; this is less than the displayed errors.

Creating these ratios we took into account the possible stellar
absorption in the H$\alpha$ line. Checking Table~\ref{t:EW_abs},
and considering the Balmer spectra of \citet{Rosa99} we suggest
that EW$_{abs}$(H$\alpha$)=6~\AA\ with a constant value along the
slit. Since EW(H$\alpha$)$\approx$15\AA\, for the emission line at
the center of \AM\,, decreasing to the galaxy's edges, this
correction is very important and without it the ratios of
[\ionn{N}{ii}]/H$\alpha$ and [\ionn{S}{ii}]/H$\alpha$ would increase from the \AM\,
centre to the edges. That could be interpreted as an increase in
of metallicity with galacto-centric distance, which is not
correct. With a measured line ratio for the central part of \AM\
($\pm$2 arcsec) $\rm ([\ionn{N}{II}]
\lambda6583/H\alpha)$=0.54$\pm$0.02, the metallicity in the center
of \AM\ is 12+log(O/H)=8.92$\pm$0.06 dex \citep{Den02} and drops
down to 8.81$\pm$0.07 in the outer parts of the galaxy. The figure
indicates that along the major axis of the \AM\ galaxy
n$_e\simeq$50~cm$^{-3}$. The measurements for detected part of
PGC~400092 give 12+log(O/H) = 8.45$\pm$0.12 dex and n$_e \simeq$
500~cm$^{-3}$.

In a similar way, we derive the gas properties along the major
axis of the ring (see Fig.~\ref{fig:AM_27_cond}). With the line
ratios measured in the central part of \AM\ ($\pm$2 central
arcsec) $\rm ([\ionn{N}{II}] \lambda6583/H\alpha)$=0.51$\pm$0.04,
the metallicity in the center of \AM\ is 12+log(O/H)=8.91$\pm$0.06
\citep{Den02}, essentially
the same value found from the major axis measurement. From the
measured [\ionn{S}{ii}] lines ratio we obtain the same value found
previously: n$_e\simeq$ 50~cm$^{-3}$.

\subsection{Newly detected H$\alpha$ emission knot}

An isolated H$\alpha$ emission knot was detected at
\mbox{$\alpha_{2000.0}$ = 19$^h$38$^m$42$^s$.7};
\mbox{$\delta_{2000.0}$ = $-$56$^{\circ}$:26':18''}, some
78\arcsec\ away from the main body of the galaxy to the NE and on
the extension of the ring's major axis. This knot is real and was
detected on all spectra observed at $\rm PA = 27\degr$ taken on
2006 September 20 and 21. The velocity distribution with distance
is shown in the top panel of Figure~\ref{fig:AM_rot_27_all}. It is
evident that the line emitting knot is fairly isolated and is very
distant from the galaxy, yet its radial velocity is close to that
of the \AM\ systemic velocity. The measured velocity for the knot
is 11645$\pm$5 km sec$^{-1}$; this is more than three standard
deviations away from the systemic velocity of \AM\ and very many
standard deviations away from the recession velocity measured for
H$\alpha$ at the NW tip of the galaxy. It is also very different
from the velocity of PGC~400092, the NW companion of \AM\,, or
from that of PGC~399718, the other companion in the triplet.

Our observations do not show a significant velocity dispersion of
the H$\alpha$ line observed from the knot, as shown in the bottom
panel of Fig.~\ref{fig:AM_rot_27_all}; a formal measurement
indicates that this H$\alpha$ line has the same FWHM ($\sim$2.4
\AA) as the reference night-sky line. The corrected FWHM$\leq$1
\AA\ for the redshifted H$\alpha$ from the knot indicates internal
motions slower than 40 km s$^{-1}$. The size of the line-emitting
region is only $\sim$5 arcsec; small, but well-resolved  by our
observations. The very weak continuum is detected; this allows a
measurement of EW(H$\alpha$) = 120$\pm$15 \AA. No additional
emission lines are visible in the spectrum.

\section{Analysis}
\label{txt:disc}

The image of the field displayed in Figure~\ref{fig:AM_direct}
shows not only \AM\, but also its two companion galaxies.
Fig.~\ref{fig:AM_direct} is a V-band image of the field obtained
with SALTICAM in the same night as the spectroscopic observations
on September 21. The image of the three galaxies allows one to
note that (a) the region around the target contains many diffuse,
low surface brightness (LSB) images that might be parts of
galaxies or LSB dwarfs at the same redshift, or distant objects in
the background, and (b) the appearance of the companion galaxy
PGC~400092 to the NW is that of a Sd galaxy with a similar overall
size to that of \AM\,. The LSB objects are also visible on
Digitized Sky Survey images of the region.

We performed unsharp masking of Figure~\ref{fig:AM_direct} to
emphasize the dust lane; this is shown in
Figure~\ref{fig:AM_direct_masked} and, contrary to the claim of
\citet{Resh06} that the dust lane is split and embraces the galaxy
nucleus from SE and NW, indicates that the lane is fairly
straight, passes south and west of the brightest part of the
galaxy, and is probably not split at all. The stars in
Fig.~\ref{fig:AM_direct_masked} have the shapes of crescent moons.
This arises from telescope optical problems which are being ironed
out during the Performance Verification process, and have been
emphasized by the unsharp masking.

The measured ratio of emission lines to corrected
H$\alpha$, and the possibly very weak [\ionn{O}{iii}]
$\lambda$5007 emission detected in our July 2006 short-wavelength
spectra, puts this object at the border between starburst nuclei
(SBN) and LINERs. Norris et al. (1990) found no compact radio core
in this galaxy and for this reason it should be classified as SBN;
this is in agreement with the previous conclusions of
\citet{Al91}.

The curves shown in Fig.~\ref{fig:AM_rot_130b} indicate that the
gas rotation along the major axis has its maximum at $\sim$240 km
sec$^{-1}$ and not at 195 km sec$^{-1}$ as given by
\citet{Resh06}, and that this maximum is reached asymptotically
for the NE part of the galaxy. Figire~\ref{fig:AM_rot_130a}
shows that our measurements are compatible with those of
\citet{Resh06} for the regions of overlap. The last points of the
rotation curve branch of the SE part of the galaxy, from
galacto-centric distance of 6\arcsec\ to 10\arcsec\,, drop from
200$\pm$7 km sec$^{-1}$ to 150$\pm$7 km sec$^{-1}$ in both
H$\alpha$ and [\ionn{N}{ii}] $\lambda$6583 lines.
This drop is gradual from 6\arcsec\ to 8\arcsec\, but shows a
step-like drop at this location, followed by a recovery with a
similar distance-velocity gradient as for the central part of the
galaxy.

A comparison of the major axis rotation curves shown in
Fig.
~\ref{fig:AM_rot_130b} shows clearly the difference between the
kinematic behaviour of the two \Na\,D absorption lines and the
H$\alpha$, [\ionn{N}{ii}] $\lambda$6583 and [\ionn{S}{ii}]
$\lambda$6716 emission lines. At this point it is worth
discussing the origin of the \Na\ absorption lines. These could be
produced by stellar photospheres, or by diffuse gas in the
interstellar medium of \AM\,. For the case of dwarf starburst
galaxies, Schwartz \& Martin (2004) used giant and supergiant
stars to show that the EW of the \Mg\ triplet near 5180\AA\, should
be twice that of the \Na\ lines. If this would be the case for
\AM\, then our blue spectrum where the \Mg\ triplet is barely
visible would rule out a major \Na\ absorption contribution from
stars.

\begin{figure*}
{\centering
 \includegraphics[clip=,angle=-90,width=17.0cm]{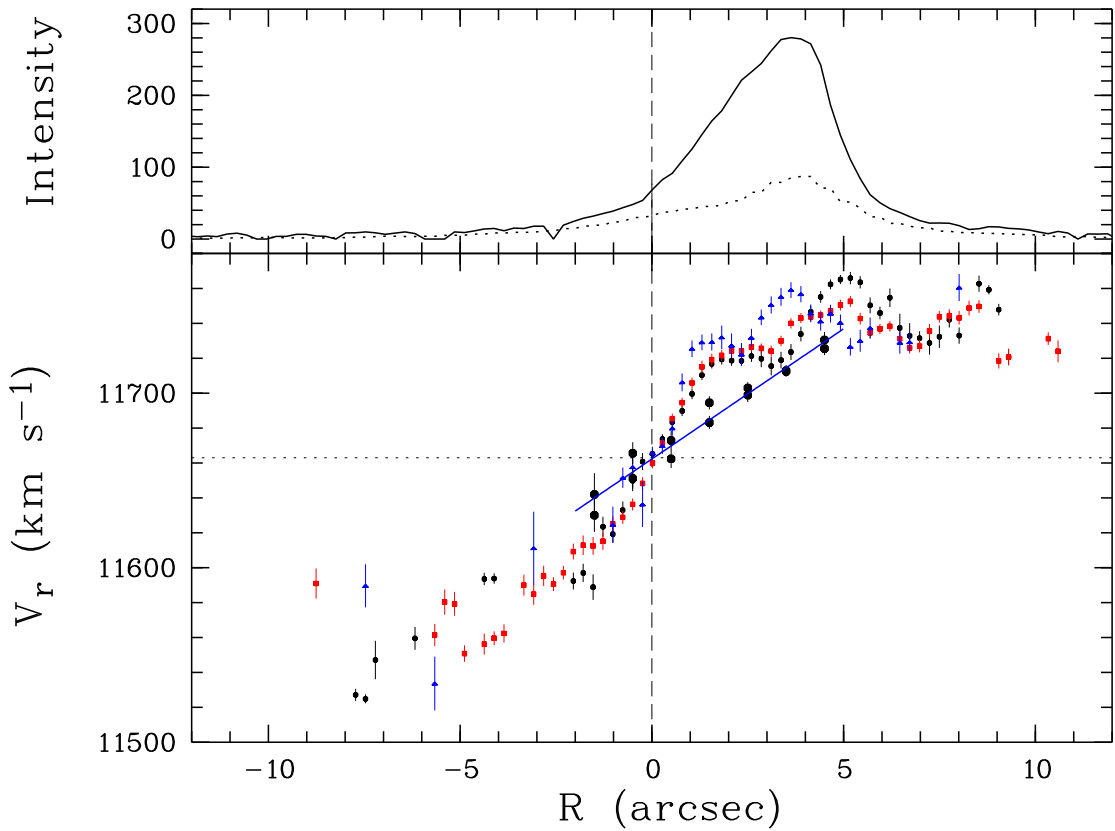}
}
 \caption{%
{\it Top panel:} The solid line shows the profile of the net H$\alpha$
flux along the slit at $\rm PA = 27\degr$\, with the continuum
subtracted. NE is to the right. The short-dashed line shows the
continuum intensity distribution along the slit and near the
H$\alpha$ line. {\it Bottom panel:} The radial velocity
distribution along the major axis of the ring of \AM\, at
PA=27$^{\circ}$. The black squares, red squares and blue triangles
represent measurements of the emission lines H$\alpha$,
[\ionn{N}{ii}] $\lambda$6583 and [\ionn{S}{ii}] $\lambda$6716
respectively. The filled black circles show the stellar velocity
distribution of the absorption doublet \Na\,D $\lambda\lambda$5890,
5896. The solid blue line is the result of a linear fit to all
measurements of the \Na\,D lines (see Section~\ref{txt:results} for
more details).
    \label{fig:AM_rot_27}}
\end{figure*}

However, in giant galaxies such as \AM\, the stellar populations
are better represented by main sequence stars. These have stronger
photospheric \Na\ than \Mg\ (e.g., a M0V star from the same library
as used by Schwartz \& Martin (2004) has EW(\Mg)=20\AA\, and
EW(\Na)=12\AA\,. While it is not possible to separate the stellar
\Na\ from the interstellar absorption, we can accept that at the
least a fraction, and perhaps all of the observed absorption
represents the stars in the galaxy. For example, in M82 Saito et
al. (1984) detected \Na\ absorption that they attributed to stars
and interpreted as solid-body rotation.

Assuming that most of the \Na\ absorption is photospheric,
this would indicate that, while the gaseous component follows a
``normal'' galactic rotation law, the stellar component rotates
almost like a solid body for $\sim$10\arcsec\ away from the
centre. The maximal rotation velocity exhibited by the stellar
component is only $\sim$150 km sec$^{-1}$ at 10\arcsec\ from the
centre for both ends of the major axis.

The extent over which the emission is observed for the ``polar
ring'' is almost the same as for the major axis, some 18\arcsec\
overall as shown in Fig.~\ref{fig:AM_rot_27}, but the derived
rotation curve is completely different. The rotation curve
indicates solid-body like rotation for 1\farcs5 to the NE (one resolution
element away from the centre, given the seeing) and for
about 5\arcsec\ to the SW. The velocity difference between the
outermost points on the slit where the absorption lines are
measured is only 90 km sec$^{-1}$. The velocity gradients shown by
the stellar components along the major axis of the galaxy and
along the axis of the PR, in regions where a linear rotation curve
can be defined, are very similar as equations (1) and (2) show. In
both cases the gradients are $\sim$19 km sec$^{-1}$ kpc$^{-1}$, where
we converted the observational gradients from equations (1) and (2)
to physical units.

\section{Interpretation}
\label{txt:interp}

At a distance to the object of 167 Mpc (H$_0$=70 km sec$^{-1}$
Mpc$^{-1}$) the radius of the galaxy to the outermost point where
emission lines are visible is $\sim$8~kpc. We found the stellar
component of a 16 kpc wide galaxy rotating as a solid body, while
its gaseous component measured at the same slit position shows a
smoothly increasing rotation curve which then flattens out. A ring
or disk feature with an extent similar to that of the galaxy is
observed at an inclination of $\sim$60$^{\circ}$ to the major axis
of the galaxy. The stellar component observed with the
spectrometer slit oriented along the major axis of the ring is
also rotating as a solid body and with a similar velocity-distance
gradient to that observed for the main body of the galaxy.

\citet{Resh06} concluded from their photometry and spectroscopy,
coupled with results of N-body modelling, that \AM\ is a PRG.
Their models indicate that the system might be the result of a
major interaction between a ''donor'' galaxy with a 17 kpc stellar
disk and a 42 kpc gaseous disk, with a total mass of
3.6$\times10^{11}$ M$_{\odot}$, which encountered a
2$\times10^{11}$ M$_{\odot}$ and 14 kpc wide ''receptor'' galaxy
some 1.6 Gyrs ago with an impact parameter of 130~kpc and a
relative velocity of 145 km sec$^{-1}$. This encounter transferred
a large quantity of matter (stars, gas, and dust) from the donor
to the receptor galaxy resulting in the formation of the polar
ring which is inclined with respect to the galaxy disk and is
warped. \citet{Resh06} suggested that the donor galaxy survived
and is PGC~399718, the southern companion in the triplet, and
argued that their suggestion is supported by the reddish (B-V)
colour of the galaxy and by its somewhat disturbed appearance.

In selecting this scenario in preference to those of minor mergers
calculated by them, or of other possible models for the formation
of ring galaxies, \citet{Resh06} relied primarily on the
morphological appearance of the galaxy. In particular, the minor
merger models rejected by \citet{Resh06} produced only
partially-open rings that were not closed, whereas the preferred
major merger model produced a ``closed and regular ring''  a few
10$^8$ years following the interaction.

Since the acceptance of the \citet{Resh06} scenario as the
explanation for the appearance of this system relies on their
interpretation that the ring is closed and regular, it is worth
examining whether the observations presented here support this
assertion.

\begin{figure}
 \begin{center}
 \includegraphics[clip=,angle=0,width=8.5cm]{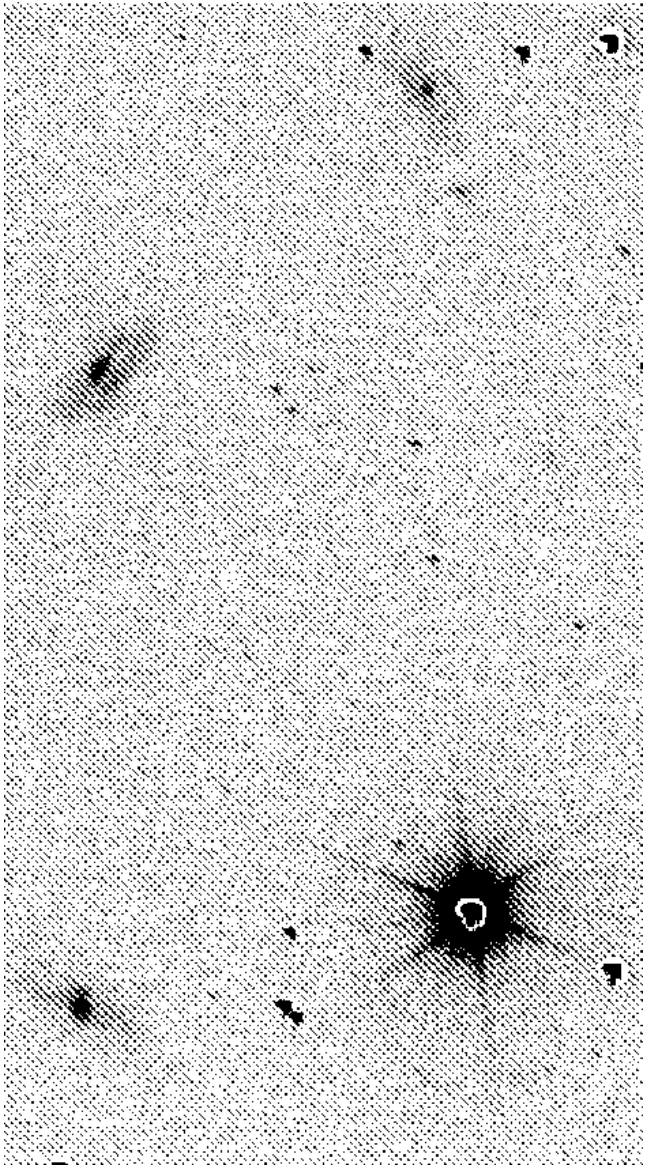}
 \end{center}
 \caption{Unsharply masked image in the V-band of \AM\, obtained with SALTICAM. This was
 cropped from Figure \ref{fig:AM_direct} to show the three galaxies and
 to emphasize the shape of the dark lane.
    \label{fig:AM_direct_masked}}
\end{figure}

The specific items resulting from
our observations that require understanding are:
\begin{enumerate}

\item   Solid-body rotation is observed for stars vs. a
''regular'' rotation for the gas at the same (projected)
locations. No differential rotation, as expected from a stellar
disk, is observed. This is true for the main body of the galaxy as
well as for the ring, though with the gas showing a different
distance-velocity gradient than the stars.

\item The ring is very faint and there is no evidence that it
contains a considerable number of stars, as would be expected from
the major merger claimed by \citet{Resh06}. Our observations of the intensity distribution
along the slit at PA=27$^{\circ}$ show that the stars producing
the continuum are located mostly where the HII is, namely some
2-5\arcsec\ NE of the centre.

\item The ring dynamics are different at its SW end, where the
line and continuum emissions are very weak and the ring is more
extended \citep{Resh06}, in comparison with the other end of the
ring.

\item The gas dynamics for the ring are very different from those
of the gas in the galaxy. Specifically, at similar extents from
the dynamical centre the gas in the ring spins much slower than
the gas in the galaxy. This, while the stellar components have
similar kinematic behaviours as evaluated from the
velocity-distance gradients.

\end{enumerate}

Apparent solid-body rotation of a galaxy could be produced, for
example, by dust extinction. Baes \etal (2003) modelled the light
propagation through a dusty galactic disk and showed that, unless
the disk is perfectly edge-on, no effects in the kinematics would
be observable. The more the disk is edge on, and the stronger the
extinction caused by the dust in the disk is, the more would the
rotation curve resemble that of a solid body. Perusal of the DSS
images of the object, of the image shown in Fig. 1 of
\citet{Resh06}, and of our Figs.~\ref{fig:AM_direct} and
\ref{fig:AM_direct_masked}, shows that \AM\ is not a purely
edge-on galaxy and that, since the disk deviation from edge-on is
definitely more than ``a few degrees'' but rather $\sim$25$^{\circ}$,
as explained below, we should not expect to
see a solid-body rotation just because of dust obscuration and
light scattering. We can, therefore, reject the possibility that
the solid-body rotation is an effect of dust obscuration.

\subsection{Stars vs. gas in the disk}

The key observation reported here is the difference in rotation
curves between the emission lines produced by the gas and the
stars as represented by the absorption lines. Such cases of
different kinematic behaviour of the gas and the stars are known
in the literature, \eg Bettoni \etal (1990), where NGC~2217 was
shown to exhibit ``counter-rotation'' in that the gas motions in
the inner parts of the galaxy indicated motions opposite those of
the stars. This was interpreted there as a consequence of a warp
in the disk coupled with the presence of a bar; this situation may
exist for \AM\ as well.

Macci{\`o} \etal (2006) tried to explain the origin of PRGs by
accretion of cold intergalactic gas. They provide in their Fig.~4
plots of simulated velocity-position diagrams for gas and stars;
the upper one, where the slit is aligned with the major axis of
the galaxy, can be compared with our Figs.~\ref{fig:AM_rot_130a}
and \ref{fig:AM_rot_130b}. It seems that the presence of a stellar
bar in \AM\ could be producing the linearly-rising stellar
rotation curve, whereas the rotation curve for the gas fits the simulation quite
well.

Since none of our observations are of photometric-quality, we rely
on parameters derived by \citet{Resh06} to characterize the
galaxy. In particular, we adopt their photometric disk parameters:
a disk exponential scale length h(B)=5".1$\pm$0".3=3.8 kpc and
their scaling to other bandpasses: h(B)/h(V)=1.18$\pm$0.11 and
h(B)/h(R)=1.25$\pm$0.12. The R-band disk scale length is,
therefore, 4.8$\pm$0.5 kpc. This is useful when comparing with
properties of other galaxies or of model galaxies.

To compare with the rotational properties of other galaxies, we
use the observations of edge-on galaxy disks from Kregel \etal
(2004) for the stellar kinematics and from Kregel \& van der Kruit
(2004) for the gas kinematics. Fig. 6 in Kregel \etal shows that
the stellar rotation curve can be almost linear with
galacto-centric distance for about 1.5 disk scale lengths and this
for galaxies earlier than Sbc. Note that this galaxy sample does
not include barred galaxies, though Kregel \etal mention that some
do show boxy or peanut-shaped bulges. The gas in none of their
galaxies (Kregel \& van der Kruit 2004) rotates with as small a
gradient with distance from the center as observed in \AM.

It is also possible to compare both the imaged galaxy and its
stellar kinematics with the diagnostic plots calculated by Bureau
\& Athanassoula (2005). Inspection of their Figs. 1 and 4
indicates that a good fit with \AM\, could be obtained for an
intermediate or strong bar viewed at least at 45$^{\circ}$ to the
bar or even edge-on, and at a disk inclination of at least
80$^{\circ}$ to the line of sight. The conclusion is that \AM\,
does probably have a fairly strong bar that is almost side-on to
our line of sight, and its disk is seen almost edge-on.

Another comparison for our rotation curve is with the collection
of template rotation curves of Catinella \etal (2006) who,
however, studied normal galaxies, not PRGs. They normalize the
rotation curves between 2 and 3 disk radii; applying this to \AM,
with the peak rotation derived from the curve, indicates that the
galaxy should have an absolute I-band magnitude brighter than --23
mag. Indeed, using the photometry from \citet{Resh06}, with a
measured M$_B \simeq$--21 mag and a color index (B-I)=2.06, the
absolute I magnitude of \AM\ is --23.06 mag. This confirms the
assumption that, in analyzing the gaseous rotation curve along the
major axis, it is a valid assumption to adopt the rotation pattern
of a regular galaxy, not that of a PRG, since the presence of the
polar ring does not affect significantly the kinematics of the
galaxy.

\subsection{HI vs. other kinematic indicators}

The HI in a number of PRGs, including \AM, was studied by van
Driel \etal (2002) with the Parkes radio telescope. This
observation produced a puzzling and troublesome result for \AM\,;
van Driel \etal reported the HI line at a heliocentric velocity of
11282$\pm$24 km sec$^{-1}$ with a full-width at half-maximum of
the two-horned profile of 193 km sec$^{-1}$. Note that their data
were taken with the Parkes multibeam system, which implies a beam
width of 14\farcm4 FWHM. The 12500 km sec$^{-1}$ bandwidth was
centered at 10000 km sec$^{-1}$ and the channel separation was 6.6
km sec$^{-1}$.

If the HI would have been associated with \AM\,, we would expect
to find the neutral hydrogen line at a similar systemic velocity
to that measured here, that in \citet{Resh01}, or that measured by
\citet{Resh06}. We would also expect a much wider HI profile than
quoted by van Driel \etal (2002), since the H$\alpha$ kinematics
indicate a width of $\sim$450 km sec$^{-1}$ along the major axis,
as befitting a major galaxy given its bright absolute magnitude of
M$_B$=--21.1 measured by \citet{Resh06}. The very wide Parkes beam
implies that all three objects were included in the measurement,
and probably many outlying HI clouds that may exist in this
neighbourhood as well, but does not explain the velocity
discrepancy since all three galaxies should have appeared on the
red shoulder of the HI profile shown by van Driel \etal

Another indication that something is wrong with the HI measurement
comes from applying the Tully-Fisher relation to \AM\,. \citet{Co06}
gives a Tully-Fisher diagram for PRGs in Fig. 2 of her
paper; these galaxies seem to follow the T-F relation for spirals
and S0 galaxies and it is worthwhile to check where \AM\ fits in
this diagram. Adopting the HI width given in van Driel \etal
(2002) indicates that \AM\ should have an M$_B\simeq$--18 mag,
completely different from the magnitude measured by
\citet{Resh06}. Adopting a velocity width as measured by us
albeit from the emission lines and not from the HI profile,
namely 450 km sec$^{-1}$, yields the proper value of
M$_B\simeq$--21 mag.

Irrespective of the explanation regarding the HI redshift
discrepancy, it is possible that extended HI is present in the
system. The possibility that such HI clouds or other gas-rich
galaxies might be present is supported by our discovery of the
H$\alpha$ knot (see below), and by the presence of a few low
surface brightness (LSB) extended objects in the immediate
vicinity. These resemble LSBs the nearby Universe that are often
found to be very gas-rich. In addition, there are a few very blue
star-like objects that stand out in comparisons of the Second
Digitized Sky Survey images in different bands.

We do not have redshifts for these LSB objects but the fact that they
are of similar sizes to the main galaxies in the \AM\ group hints
that they might be group members; such companions are seen in
other groups as well (\eg Grossi \etal 2007) and could have
interacted with \AM\ in the past. We predict that once HI
synthesis observations will be obtained for \AM\ and its
neighbours, for example with the ATNF, at least some of these
candidates and in particular the H$\alpha$ knot discovered by us
will prove to be actually gas-rich members of this group.

\subsection{The H$\alpha$ knot}

The H$\alpha$ knot reported above, which is $\sim$78 arcsec away
to the NE from the galaxy center but almost at the same velocity,
is in reality $\sim$630 kpc away in projected distance. Its
detectable H$\alpha$ emission, combined with a lack of
[\ionn{N}{ii}], [\ionn{S}{ii}] and only weak continuum emissions,
argue that this is probably a metal-poor dwarf galaxy that belongs
to the same group as \AM\,. Such objects are known as ''HII
galaxies'' (Sargent \& Searle 1970) since they show an HII region
spectrum with negligible continuum and have considerable
redshifts.

Our fitting procedure to the emission lines, used for the galaxies
and for the ring, allows the derivation of an upper limit for the
[\ionn{N}{ii}] $\lambda$6583 flux that can be used to obtain an
upper limit to the metal abundance. With a measured upper limit
line ratio of $\rm log([\ionn{N}{ii}] \lambda6583/H\alpha) =
-1.46$ the metallicity upper limit is 12+log(O/H) $<$ 8.05
\citep{Den02}. The knot appears to be somewhat metal-poor, though
we cannot set a definite upper limit on its metal abundance, and
we have shown that it also shows a very low internal velocity
dispersion as befits a dwarf galaxy.

Brosch et al. (2006) identified a considerable number of
H$\alpha$-emitting knots in the neighbourhoods of a few
star-forming galaxies qualified as ``dwarfs'' (M$_B\geq$--18) and
located in some very under-dense regions of the nearby Universe.
The study revealed these prospective neighbour galaxies through
the presence of H$\alpha$ emission at or near the central galaxy
redshift. It is possible that the knot found here is a similar
type of object.

\subsection{Differences between the two slit positions}

Spectroscopy of the ring in NGC 4650A has been reported by Swaters
\& Rubin (2003). They found a ring rotation curve that seems to
flatten out from the center to the North, but which is steadily
increasing from the center to 20 arcsec on the South side and then
flattens out. This is more pronounced for the stellar component of
the ring than for its gaseous component. The galaxy itself, an S0
as most PRGs are, shows solid-body-like stellar rotation from the
center to $\sim$15 arcsec out while the emission lines show a
different pattern of constant velocity. This they interpret as due
to the galaxy being devoid of gas while the line emission is
produced only in the ring. Comparisons of the rotational
properties of polar rings and of galaxy disks are valuable in
understanding PRGs.

We return now to the appearance of the stellar rotation curves
observed at $\rm PA = 140\degr$ and at $\rm PA = 27\degr$. These
curves, derived from the \Na\ absorption lines, are very similar.
They appear linear for a considerable distance and their
velocity-distance gradients are $\sim$19 km sec$^{-1}$ kpc$^{-1}$.

We point the reader back to Fig.~\ref{fig:AM_2D_27} where the
extent of the continuum that allows the detection and measurement
of the \Na\,D lines is considerably narrower than that for $\rm PA
= 140\degr$. The discrepancy could be resolved by assuming that
the absorption lines, and therefore most of the continuum, would
not be produced by stars in the ring, as implicitly assumed in the
previous sections, but by stars in the main galaxy, perhaps in a
stellar disk or in a strong bar. The \AM\ inclination can be
derived from the axial ratio of the galaxy given in \citet{Resh06}: i$\simeq 63
^{\circ}$.

In this case, the angle difference between the two slit positions,
$\rm 67\degr$, would explain the difference in the extent of the
linear rotation curves at the two position angles as a combination
of foreshortening and obscuration by the dust lane. The dust lane
produces about one magnitude of extinction, as the intensity
profiles along the slit in Fig. 4c of \citet{Resh06} show. The
sudden disappearance of the absorption lines only 1\farcs5 SW of the
dynamical center could be explained by the crossing of the dark
lane by the slit at this position angle. The weak intensity of the
underlying continuum of the H$\alpha$ line, plotted with a
short-dashed line in the top panel of Fig.~\ref{fig:AM_rot_27},
supports this interpretation.

The ring would, in this case, be composed mostly of gas, would be
located between us and the disk with its dark lane, and would
necessarily be much less massive than assumed by \citet{Resh06}.
The emission lines measured within
$\pm$5\arcsec\ of the kinematic centre (see e.g.,
Fig~\ref{fig:AM_rot_27}) would then be produced primarily in the
disk, while those for $\rm PA = 27\degr$ but measured at a
galacto-centric distance of more than 6\farcs5 would originate in the
ring.

The arguments presented above indicate that the model proposed by
\citet{Resh06} to explain \AM\ as a major merger, with the donor
galaxy being PGC~399718, might not fit the observations. We
therefore propose another alternative, that the unsettled disk or
ring around the galaxy was formed by accretion of cold gas from a
cosmic filament, one of the possibilities accounting for ring
galaxy formation put forward by \citet{Co06}. The presence of
anomalous HI redshifts in the region, our discovery of an apparent
dwarf HII galaxy in the group, and the circumstantial detection of
large but low surface brightness galaxies in the immediate
vicinity of \AM\,, albeit lacking redshifts at present, argue in
favour of this interpretation.

Perusing the large-scale structures identified in this region by
Radburn-Smith \etal (2006), specifically those in Panel 6 of their
Figure 4, indicates that the location of \AM, at $\rm l \simeq 341.02$,
$\rm b\simeq -28.73$, corresponds to the tip of a galaxy filament
extending out of the Zone of Avoidance. This might be a distant
structure related to the Centaurus wall and the Norma and
 Pavo II clusters of galaxies at lower redshifts,
through which intergalactic matter is accreted by the galaxy and
forms the ring.

Models of cold gas accretion from cosmic filaments by Macci{\`o}
\etal (2006) show how a ring galaxy, such as NGC 4650A or for that
matter \AM\, could be formed by such a process. Their simulations
show that the accreted gas is not completely cold but rather at
15,000K due to its collapse within the gravitational potential of
the filamentary structure. Moreover, they mention that some of the
gas might also be shock-heated by the halo potential.

A similar process could take place in \AM\,. There is no clear-cut
evidence that the ring is closed or relaxed, or that it has a
substantial stellar component. Its disturbed appearance at its SW
end is more similar to that of an assemblage of diffuse gas
clouds, not of a coherent and relaxed structure. The NE part is
smaller and sharper; it is possible that accreted gas collides
there with itself, becomes compressed and shocked, and reaches
higher temperatures that produce the enhanced line emission. At
this location the accreted gas could perhaps enter a circular or
quasi-circular orbit.

An alternative could be that in the \AM\ case we are indeed
witnessing a merger with a gas-rich galaxy, which takes place in a
polar configuration. This is, in a way, similar to the major
merger scenario of \citet{Resh06} with the exception that the
''donor'' galaxy would now be the ring itself. The argument
reducing the likelihood of this explanation is the lack of a
significant stellar continuum from the ring, indicating its
low mass.

\subsection{The shape of the dark matter halo of \AM\ }

Considering the two gas rotation curves, the one along the
galaxy's major axis and the other along the ring's major axis, one
observation is in order. The two rotation curves derived from the
emission lines extend a similar distance from the galaxy's
kinematic centre, are presumably in the same dark matter potential
well if \AM\ is indeed a PRG, yet show a completely different full
amplitude. While the galaxy major axis rotation curve has a full
end-to-end amplitude of $\sim$450 km sec$^{-1}$, that for the ring
has a full amplitude of only $\sim$240 km sec$^{-1}$. The
asymptotic rotation of the ring is slower than the asymptotic
rotation of the galaxy.

The formation of PRGs has been studied by Bournaud \& Combes
(2003) via N-body simulations. They discussed, in particular,
cases when both the galaxy disk and the ring contain gas. Their
argument was that in such cases the polar ring must, by necessity,
be wider than the galaxy. If this is not the case, the gas in the
ring would interact with the gas in the disk and one of the
components would join the other. Two orthogonal, or almost
orthogonal gas rings, can coexist in the same galaxy only if they
have different radii and do not cross each other. Such crossing
presumably occurs in NGC 660, where both the disk and the ring
contain gas; the N660 system is unstable and according to Bournaud
\& Combes did not have sufficient time to dissolve the ring since
its formation.

The specific question of the DM halo shape in PRGs was studied by
Iodice \etal (2003). They explained that the ring material would
move slower than the gas in the disk if the gravitational
potential would be oblate, like the flattened disk galaxy. In this
case the ring would be elliptical and would show a lower observed
velocity than the disk at its outermost locations (see their Fig.
3). In the case of \AM\,, since the ring and the galaxy appear to
have the same size but the ring must be wider in order to avoid
crossing the disk, a possible conclusion would be that the ring is
elliptical with its major axis close to our line of sight to the
object and its minor axis seen almost perpendicular to the disk.
This way, the ring could indeed be larger than the galaxy, the gas
in the ring and that in the galaxy would not cross, and the
velocities at the apo-galactic ring locations would be slower than
in the galaxy disk. In this case the outermost visible ring
segments would correspond to locations near the ends of the minor
axis and ring material should show there its highest orbital
speed, larger than that of the galactic disk. As this is not
observed, we conclude that the disk and the ring in \AM\ are of
similar sizes, their contents do cross, and the system is
unstable.

With the additional kinematic information now available, \AM\
could be considered a test case for DM gravitational potential
tracing.
The discussion of PRGs by \citet{Co06} was based on the hope that
PRGs would prove to be useful probes of the DM potential in which
a galaxy and its polar ring find themselves. \citet{Co06} found that the
rings in observed PRGs show faster rotation than the maximal
velocity observed in the host galaxy. The theoretical prediction
is in the opposite direction to the observations, namely rings in
PRGs devoid of DM halos or with spherical halos should be rotating
slower than their galaxies. According to \citet{Co06}, this effect
should be accentuated for flattened or oblate DM halos.

We find that \AM\ fulfills the theoretical predictions for
non-spherical oblate haloes in that the polar ring does
rotate slower than its host galaxy. Any DM halo of \AM\,, if it
exists at all, would have to be flattened along the barred disk,
but this configuration could not be stable on the long run because
the ring would cross the disk. Resolving this possibility and
deriving more constraints on the existence and shape of a possible
DM halo for \AM\ would require detailed modelling and further
observations that are not within the scope of this paper.

\section{Conclusions}
\label{txt:summ}

We presented observations obtained with SALT and RSS during their
performance verification phase that emphasize the long-slit
capabilities of the RSS for galaxy observations. We traced the
stellar and gaseous rotation curves for the major axis of the
galaxy and for the major axis of a polar ring-like feature almost
perpendicular to the disk of the galaxy. We showed that, while the
gas rotates regularly when sampled along the galaxy major axis,
the stellar component shows rotation like a solid body, supporting
an interpretation that this is an object with a strong bar viewed
almost side-on.

The ionized gas rotation along the major axis of the ring was
found to be much less regular than along the major axis of the
galaxy and shows a somewhat shallower gradient with
galacto-centric distance. The \Na\ stellar rotation from the
$\sim$6.5 kpc ring segment where the lines are measurable shows a
similar distance dependence to that seen along the galaxy's major
axis. The systemic velocity derived by us for \AM\ differs from
previously published values. We propose that the discrepancy of
rotation curves along the two position angles can be resolved by
recognizing that the absorption lines are probably produced only
by the main galaxy or its bar, and not by the ring where only
emission lines are produced.

We discovered a small H$\alpha$ knot at a projected distance of
about 700 kpc from \AM\, but at a similar velocity, which we
interpret as a fourth member of this compact group of galaxies,
presumably a metal-poor dwarf galaxy. The lack of continuum
emission for this object while only the H$\alpha$ line is detected
indicates that it might be forming stars for the first time. The
low velocity dispersion measured from the knot indicates its low
mass.

We argue that a more plausible explanation to the major merger
scenario proposed by \citet{Resh06} to explain \AM\ could be the
slow accretion of cold cosmic gas along a galaxy filament directed
to the \AM\ region.
In the cold gas accretion case the flow is probably towards the
galaxy from the South-West and becomes more compressed at the NE
end of the polar ring feature. We point out that the kinematic
properties we measured follow the theoretical predictions for PRGs
in a dark matter halo that is not spherical, but is flattened
along the plane of the galaxy.

\section*{Acknowledgments}
This paper was written while NB was a sabbatical visitor at the
South African Astronomical Observatory in Cape Town; NB is
grateful for this opportunity offered by the SAAO management. We
are grateful for the generous allocation of SALT observing time
during the PV phase to complete this project. We acknowledge a
private communication from Vladimir P. Reshetnikov concerning this
galaxy. We acknowledge the use of products of the second Digitized
Sky Survey produced at the Space Telescope Science Institute under
U.S. Government grant NAG W-2166. The images are based on
photographic data obtained using the UK Schmidt Telescope. The UK
Schmidt Telescope was operated by the Royal Observatory Edinburgh,
with funding from the UK Science and Engineering Research Council
(later the UK Particle Physics and Astronomy Research Council),
until 1988 June, and thereafter by the Anglo-Australian
Observatory. The blue plates of the southern Sky Atlas and its
Equatorial Extension (together known as the SERC-J), as well as
the Equatorial Red (ER), and the Second Epoch [red] Survey (SES)
were all taken with the UK Schmidt. An anonymous referee
provided some insightful comments that improved the clarity of the
presentation.

\bsp

\label{lastpage}


\begin{thebibliography}{99}

\bibitem[\protect\citeauthoryear{Allen et al.}{1991}]{Al91}
    Allen at al. 1991, \mnras, 248, 528

\bibitem[\protect\citeauthoryear{Baes et al.}{2003}]{2003MNRAS.343.1081B}
    Baes M., et al., 2003, MNRAS, 343, 1081

\bibitem[\protect\citeauthoryear{Bekki}{1998}]{1998ApJ...499..635B}
    Bekki K., 1998, ApJ, 499, 635

\bibitem[\protect\citeauthoryear{Bettoni, Fasano, \& Galletta}{1990}]{1990AJ.....99.1789B}
    Bettoni D., Fasano G., Galletta G., 1990, AJ, 99, 1789

\bibitem[\protect\citeauthoryear{Bournaud \& Combes}{2003}]{2003A&A...401..817B}
    Bournaud F., Combes F., 2003, \aap, 401, 817


\bibitem[\protect\citeauthoryear{Brosch et al.}{2006}]{BBM06}
    Brosch, N., Bar-Or, C., and Malka, D. 2006, \mnras, 368, 864

\bibitem [\protect\citeauthoryear{Brosch}{1985}]{Brosch85} Brosch, N. 1985, A\&A, 153, 199

\bibitem [\protect\citeauthoryear{Brosch}{1987}]{Brosch87} Brosch, N. 1987, Mercury, 16, 174

\bibitem[\protect\citeauthoryear{Buckley, Swart, \&
Meiring}{2006}]{2006SPIE.6267E..32B} Buckley D.~A.~H., Swart
G.~P., Meiring J.~G., 2006, SPIE, 6267,

\bibitem[\protect\citeauthoryear{Bureau \&
Athanassoula}{2005}]{2005ApJ...626..159B} Bureau M., Athanassoula
E., 2005, ApJ, 626, 159

\bibitem[\protect\citeauthoryear{Burgh et al.}{2003}]{2003SPIE.4841.1463B}
Burgh E.~B., Nordsieck K.~H., Kobulnicky H.~A., Williams T.~B.,
O'Donoghue D., Smith M.~P., Percival J.~W., 2003, SPIE, 4841, 1463

\bibitem[\protect\citeauthoryear{Combes}{2006}]{Co06} Combes, F. 2006, EAS, 20, 97

\bibitem[\protect\citeauthoryear{Catinella, Giovanelli, \&
Haynes}{2006}]{2006ApJ...640..751C} Catinella B., Giovanelli R.,
Haynes M.~P., 2006, ApJ, 640, 751

\bibitem[\protect\citeauthoryear{Denicol\'o et al.}{2002}]{Den02}
    Denicol\'o, G., Terlevich, R., \& Terlevich, E. 2002, \mnras, 330, 69

\bibitem[\protect\citeauthoryear{Gonz\'{a}lez-Delgado et al.}{1999}]{Rosa99}
    Gonz\'{a}lez-Delgado, R.M., Leitherer, C., \& Heckman, T.M. 1999, \apjs, 125, 489

\bibitem[\protect\citeauthoryear{Grossi et al.}{2007}]{2007MNRAS.374..107G}
Grossi M., Disney M.~J., Pritzl B.~J., Knezek P.~M., Gallagher
J.~S., Minchin R.~F., Freeman K.~C., 2007, MNRAS, 374, 107

\bibitem[\protect\citeauthoryear{Kregel \& van der
Kruit}{2004}]{2004MNRAS.352..787K} Kregel M., van der Kruit P.~C.,
2004, MNRAS, 352, 787

\bibitem[\protect\citeauthoryear{Kregel, van der Kruit, \&
Freeman}{2004}]{2004MNRAS.351.1247K} Kregel M., van der Kruit
P.~C., Freeman K.~C., 2004, MNRAS, 351, 1247

\bibitem [\protect\citeauthoryear{Hagen-Thorn et al.}{2005}]{HTetal05} Hagen-Thorn, V.A., Shalyapina, L.V.,
      Karataeva, G.M., Yakovleva, V.A., Moiseev, A.V., and Burenkov, A.N.
      2005, Astron. Reports, 49, 958

\bibitem[\protect\citeauthoryear{Iodice et al.}{2003}]{2003ApJ...585..730I}
Iodice E., Arnaboldi M., Bournaud F., Combes F., Sparke L.~S., van Driel
W., Capaccioli M., 2003, ApJ, 585, 730

\bibitem [\protect\citeauthoryear{Karataeva et al.}{2004}]{Karetal04} Karataeva, G.M., Tikhonov, N.A.,
      Galazutdinova, O.A.,
      Hagen-Thorn, V.A., and Yakovleva, V.A. 2004, A\&A, 421, 833

\bibitem[\protect\citeauthoryear{Kobulnicky et
al.}{2003}]{2003SPIE.4841.1634K} Kobulnicky H.~A., Nordsieck K.~H., Burgh
E.~B., Smith M.~P., Percival J.~W., Williams T.~B., O'Donoghue D., 2003,
SPIE, 4841, 1634

\bibitem[\protect\citeauthoryear{Kniazev et al.}{2004}]{SHOC} Kniazev, A.Y., Pustilnik, S.A.,
   Grebel, E.K., Lee, H., \& Pramskij, A.G. 2004, ApJS, 153, 429

\bibitem[\protect\citeauthoryear{Kniazev et al.}{2005}]{Sextans} Kniazev, A.Y., Grebel, E.K.,
   Pustilnik, S.A., Pramskij, A.G., \& Zucker, D. 2005, \aj, 130, 1558

\bibitem[\protect\citeauthoryear{Iodice et al.}{2002}]{2002AJ....123..195I}
Iodice E., Arnaboldi M., De Lucia G., Gallagher J.~S., III, Sparke
L.~S., Freeman K.~C., 2002, AJ, 123, 195

\bibitem[\protect\citeauthoryear{Iodice et al.}{2006}]{2006ApJ...643..200I}
Iodice E., et al., 2006, ApJ, 643, 200

\bibitem[\protect\citeauthoryear{Leitherer}{1999}]{Leitherer99} Leitherer, C. \etal 1999, ApJS, 123, 3

\bibitem[\protect\citeauthoryear{Lynds \& Toomre}{1976}]{LT76} Lynds, R. \& Toomre, A. 1976, ApJ, 209, 382

\bibitem[\protect\citeauthoryear{Macci{\`o}, Moore, \&
Stadel}{2006}]{2006ApJ...636L..25M} Macci{\`o} A.~V., Moore B.,
Stadel J., 2006, ApJ, 636, L25

\bibitem[\protect\citeauthoryear{Mayya \& Korchagin}{2001}]{MK01} Mayya, Y.D. \& Korchagin, V.
     2001, ApSSS, 277, 339 (on-line  revision in 2006)

\bibitem [\protect\citeauthoryear{Mazucca et al.}{2001}]{Mazetal01} Mazucca, L.M., Knapen, J.H., Regan, M.W. \& B\"{o}ker, T.
 2001, in {\it The Central kiloparsec of Starbursts and AGN}, J.H.
 Knapen \etal, eds. San Francisco: ASP, 573

\bibitem[\protect\citeauthoryear{Norris et al.}{1990}]{1990ApJ...359..291N}
Norris R.~P., Kesteven M.~J., Troup E.~R., Allen D.~A., Sramek
R.~A., 1990, ApJ, 359, 291

\bibitem[\protect\citeauthoryear{O'Donoghue et
al.}{2006}]{2006MNRAS.372..151O} O'Donoghue D., et al., 2006,
MNRAS, 372, 151

\bibitem[\protect\citeauthoryear{Persic, Salucci, \&
Stel}{1996}]{1996MNRAS.281...27P} Persic M., Salucci P., Stel F.,
1996, MNRAS, 281, 27 (erratum 1996, MNRAS, 283, 1102)

\bibitem[\protect\citeauthoryear{Radburn-Smith et
al.}{2006}]{2006MNRAS.369.1131R} Radburn-Smith D.~J., Lucey J.~R.,
Woudt P.~A., Kraan-Korteweg R.~C., Watson F.~G., 2006, MNRAS, 369,
1131

\bibitem[\protect\citeauthoryear{Reshetnikov et al.}{2001}]{Resh01}
Reshetnikov V.~P., Fa{\'u}ndez-Abans M., de Oliveira-Abans M., 2001, \mnras, 322, 689

\bibitem[\protect\citeauthoryear{Reshetnikov}{2004}]{Resh04} Reshetnikov, V.P. 2004, A\&A, 416, 889

\bibitem[\protect\citeauthoryear{Reshetnikov et al.}{2006}]{Resh06}
Reshetnikov, V., Bournaud, F., Combes, F., Fa{\'u}ndez-Abans, M., and de
Oliveira-Abans, M.: 2006, {\it Astron. Astroph.} 446, 447.

\bibitem[\protect\citeauthoryear{Sargent \&
Searle}{1970}]{1970ApJ...162L.155S} Sargent W.~L.~W., Searle L.,
1970, ApJ, 162, L155

\bibitem[\protect\citeauthoryear{Schweizer et al.}{1987}]{Schwetal87} Schweizer, F., Ford, W.K. Jr.,
     Jederzejewsky, R. \& Giovanelli, R. 1987, ApJ, 320, 454

\bibitem[\protect\citeauthoryear{Shane}{1980}]{Shane80} Shane, W.W. 1980, A\&A 82, 314


\bibitem[\protect\citeauthoryear{Swaters \&
Rubin}{2003}]{2003ApJ...587L..23S} Swaters R.~A., Rubin V.~C.,
2003, ApJ, 587, L23

\bibitem[\protect\citeauthoryear{van Driel et
al.}{2002}]{2002A&A...386..140V} van Driel W., Combes F.,
Arnaboldi M., Sparke L.~S., 2002, A\&A, 386, 140

\bibitem[\protect\citeauthoryear{Whitmore et al.}{1990}]{Whietal90}
Whitmore B.~C., Lucas R.~A.,
McElroy D.~B., Steiman-Cameron T.~Y., Sackett P.~D., Olling R.~P.,
1990, AJ, 100, 1489


\bibitem[\protect\citeauthoryear{Zasov et al.}{2000}]{Zasov00}  Zasov, A., Kniazev, A., Pustilnik, S., et al.
     2000, A\&AS, 144, 429

     \bibitem[\protect\citeauthoryear{Woudt et al.}{2006}]{2006MNRAS.371.1497W}
Woudt P.~A., et al., 2006, MNRAS, 371, 1497

\end{thebibliography}
\end{document}